\documentclass[12pt]{amsart}
\usepackage{setspace,graphicx,srcltx,enumitem,harvard,subfigure}
\usepackage[foot]{amsaddr}
\usepackage{rotating, lscape}
\usepackage{bm}
\usepackage{amssymb,amsmath,amsfonts,amsthm}
\usepackage{booktabs,colortbl,multirow}
\usepackage{upgreek}
\usepackage{enumitem}
\usepackage{adjustbox}
\usepackage{graphicx} 
\usepackage{geometry}
\usepackage{tikz}
\usepackage{setspace}
\usepackage{graphicx}
\usepackage{float}
\usepackage{comment}
\usepackage{url}
\usepackage{caption}
\usepackage{ulem}

\usepackage{graphicx} %package to manage images
\graphicspath{ {./images/} }

\usepackage{xcolor}
\usepackage{colortbl}
\usepackage{booktabs}
\usepackage{rotating}
\usepackage{tabularx}
\usepackage{adjustbox}
\usepackage{tablefootnote}
\usepackage{longtable}
\usepackage[flushleft]{threeparttable}

\allowdisplaybreaks
\usepackage{ragged2e}
\def\0{{\bm{0}}}

\def\be{\begin{eqnarray}}
\def\ee{\end{eqnarray}}

\thanks{Julia Hatamyar, Centre for Health Economics, University of York. Christopher F. Parmeter, Department of Economics, University of Miami, Coral Gables, FL 33146; Corresponding Author e-mail: julia.hatamyar@york.ac.uk All {\tt R} and {\tt Stata} code used in this paper is available upon request.}
\IfFileExists{upquote.sty}{\usepackage{upquote}}{}

\begin{document}
	
% \title{Some \textit{Skewed} Results on the Stochastic Frontier Model}
\title[COVID and Evictions]{Local Eviction Moratoria and the Spread of COVID-19}
\date{\today}

\author{Julia Hatamyar}
\author{Christopher F. Parmeter}

\thanks{We thank participants at the University of York Applied Microeconomics Cluster Seminar and the University of Miami for their invaluable feedback. The usual disclaimer applies.}

\begin{abstract}
At various stages during the initial onset of the COVID-19 pandemic, various US states and local municipalities enacted eviction moratoria. One of the main aims of these moratoria was to slow the spread of COVID-19 infections. We deploy a semiparametric difference-in-differences approach with an event study specification to test whether the lifting of these local moratoria led to an increase in COVID-19 cases and deaths. Our main findings, across a range of specifications, are inconclusive regarding the impact of the moratoria - especially after accounting for the number of actual evictions and conducting the analysis at the county level. We argue that recently developed augmented synthetic control (ASCM) methods are more appropriate in this setting. Our ASCM results also suggest that the lifting of eviction moratoria had little to no impact on COVID-19 cases and deaths. Thus, it seems that eviction moratoria had little to no robust effect on reducing the spread of COVID-19 throwing into question its use as a non-pharmaceutical intervention.
\end{abstract}

\maketitle

%Homelessness leads to mental health issues. Put into conclusions that our paper is only focused on moratoria and COVID, not other perhaps more germane issues related to homelessness. 

\section{Introduction}

With the near universal shutdown of the U.S. economy following the outbreak of COVID-19, many individuals could not (or chose not to) work, which led to concerns over late rental payments. To combat these concerns many states and local municipalities enacted (at various times) eviction moratoria that prevented \textit{qualified} renters from being evicted. One of the primary motivations for these moratoria was to help prevent the spread of COVID-19 given the tangible public health risks of evicting people while a highly contagious respiratory disease was spreading.\footnote{https://www.vox.com/21569601/eviction-moratorium-cdc-covid-19-congress-rental-assistance-rent-crisis.}

%On September 4th, 2020 the US Center for Disease Control (CDC), invoking its authority under the Public Health Service Act,\footnote{Pub.L. 78–410, 1944} passed a broad eviction moratorium across the country running through June 30th, 2021, due to the expiration of the partial eviction moratorium from the CARES Act.\footnote{H.R. 748} Prior to this, 44 US states also enacted their own local eviction moratoria at different times.  Specifically, the CDC Act aimed to mitigate the ``spread of COVID-19 within congregate or shared living settings, or through unsheltered homelessness'' \cite{centers2021temporary}. 

On the surface, eviction moratoria seem a prudent policy measure. However, given a raft of other COVID-19 policies that were already in place across most US states, the efficacy of such a policy with respect to preventing the spread of COVID-19 is not obvious.\footnote{In addition to slowing/mitigating the spread of COVID-19 due to evictions, the moratorium kept tenants in their homes at a time when unemployment was high due to economy-wide impacts from the pandemic.} This suggests that identification of such an impact is likely to prove difficult. This is succinctly characterized by \citeasnoun[pg. 154]{GOODMAN-BACON_MARCUS:2020}: ``Good control groups will have to match treatment groups on many dimensions. Smart research designs will try to focus on situations where treatment and control groups differ only by the introduction of a single COVID policy (or, at least, only few policies).'' 

To date the findings in the literature related to the ability of eviction moratoria to slow the spread of COVID-19 are mixed \citeaffixed{GOODMAN-BACON_MARCUS:2020}{as presaged by}. The first attempt to study the impact of eviction moratoria on the spread of COVID-19 is \citeasnoun**{LEIFHEIT_ETAL:2021} who use data from the 44 states that ever instituted an eviction moratoria from the period March 13 to September 3, 2020. \citeasnoun**{LEIFHEIT_ETAL:2021} deploy a difference-in-difference (DiD) approach with a two-way fixed effects event-study specification and find that both COVID-19 incidence and mortality increased steadily in states \textbf{after} the moratoria expired. They find that a spike in deaths due to evictions occurring after expiration of moratoriums \textit{preceded} a spike in cases, which occurred almost 10 weeks later. In related work, \citeasnoun**{NANDE_ETAL:2021}, use a simulated model of viral transmissions, and predict that evictions increase COVID-19 infection risk. They then apply their simulated model to Philadelphia using locally-specific parameters, and conclude that eviction moratoria are an effective and important policy measure. 

Using a panel of individuals who were diagnosed with COVID-19 and a Cox DiD regression, \citeasnoun**{sandoval2021eviction} find an increased likelihood of a COVID diagnosis after state-level moratoria were lifted. \citeasnoun**{jowers2021housing} study the impact of ``housing precarity policies" at the county level, which include both eviction and utility disconnection moratoria, on added COVID-19 cases and deaths, using a traditional panel fixed effects regression. Although the authors find that eviction moratoria reduce infections and deaths by a significant amount, their econometric model raises causal identification concerns - and does not control for any other local policies in place. In contrast to the above studies, \citeasnoun**{pan2020covid} examine a variety of non-pharmaceutical interventions (including eviction moratoria) using a negative binomial specification, and do \textit{not} find any statistically significant impact of eviction policies on COVID-19 spread.\footnote{The authors find that only shelter-in-place, stay at home measures, mask mandates, and travel restrictions achieved a significant effect.}  

Our work here critically examines the impact of local eviction moratoria on COVID-19 incidence and mortality. Although the work of \citeasnoun**{LEIFHEIT_ETAL:2021} and \citeasnoun**{sandoval2021eviction} are crucially important for understanding the potential causal effects of the state level eviction moratoria on limiting the spread of the COVID-19 virus, we nonetheless demonstrate that their results are not robust when replicated using alternative econometric techniques. This paper also differs from previous work in that we include actual eviction numbers as a control, perform analysis at the county level, and focus mainly on large metropolitan centers (where population density is increased). 

%We begin by mimicking the research of \citeasnoun**{LEIFHEIT_ETAL:2021}. This re-examination is important as \citeasnoun[pg. 783]{PENG_ETAL:2006} note that ``Scientific evidence is strengthened when important findings are replicated by multiple independent investigators using independent data, analytical methods, laboratories, and instruments.'' 

We preview our results here. First, %in the spirit of \citeasnoun{PENG_ETAL:2006}, 
we construct a dataset mimicking that of \citeasnoun**{LEIFHEIT_ETAL:2021}. We also buttress this exercise with several other extensions which we believe lend credence to the estimation of a causal effect, and fail to find that expiring eviction moratoria had quantitatively meaningful impacts on either cases or deaths.\footnote{Replication details and results can be found in the appendix.} Next, we construct a new dataset at the county level, for a variety of metropolitan areas. We use Princeton Eviction Lab \cite{evictionlab} data on the actual number of evictions in each of these counties by week, which allows us to control for this important confounding variable. Lastly, we repeat the analysis using three different estimators (each of which has merits beyond the simple two-way fixed effects DiD approach), and again fail to find significant evidence that expiring moratoria had any causal impact on either cases of, or deaths from, COVID-19.  

One reason that we believe the main finding of \citeasnoun{LEIFHEIT_ETAL:2021} dissipates is that the timing differences of expiring eviction moratoria suggest that an alternative weighting scheme be used \cite**{GOODMAN-BACON_MARCUS:2020,SUN_ABRAHAM:2020,de2020two,borusyak2021revisiting,baker2022much}. This scheme weights the treatment effects based on the cohorts of time from the expiration of the moratoria which has meaningful consequences not only for the estimates, but also the standard errors.\footnote{These alternative methods are also in alignment with the recommendations of \citeasnoun{GOODMAN-BACON_MARCUS:2020}} When using more recent statistical models to account for this requirement, the \citeasnoun{LEIFHEIT_ETAL:2021} analysis fails at the state level. However, even if the results did hold, the county level is arguably the more relevant geographic area of analysis due to significant differences between state and county-level policy implementation (for example Austin's local moratoria in contrast to the lack of a binding Texas order). Finally, although \citeasnoun{LEIFHEIT_ETAL:2021} do control for various policies and population size in their specifications, they do not control for political or eviction-related potential confounders. These variables are likely to impact both the implementation of eviction laws and the number of COVID-19 cases and deaths. 

Lastly, even with the cohort specific weighting, we argue that the most appropriate method to study the potential causal impact of eviction moratoria on the transmission of COVID-19 is augmented synthetic control (ASC) with staggered adoption \cite{BEN-MICHAEL_ETAL:2021}. This method constructs synthetic control observations that can be compared to the treated group while accounting for the staggered adoption that is prevalent in many event study applications. It is an ideal tool since even taking out county-specific averages, as done in a DiD, is unlikely to be credible given the substantial heterogeneity that is likely to be present in differences between counties, both in trends and in levels. As \citeasnoun[pg. 2561]{IMBENS:2022} notes ``The basic synthetic control method \dots has in a short time found many applications in a wide range of fields, including \dots the effects of country- or state-level COVID-19 policies.'' Again, using ASC with staggered adoption, our findings remain consistent. Once the moratoria expires, there is no statistically significant effect on COVID-19 cases or deaths. 

Overall, our main finding is that while eviction moratoria certainly helped to keep people in their homes during a time of significant economic upheaval, the moratoria themselves had no statistically significant effect on COVID transmission. The fact that our findings differ from most previous work is likely due to the inability of studies at the state level to pinpoint specific transmission patterns that are likely to vary at a local scale, other policy devices already in place prior to any moratoria expiring, individuals being aware of the transmission of COVID and taking necessary steps to avoid infection, and eviction moratoria not being truly complete bans on evictions. All of these issues combined make it plausible that an eviction moratoria, as a policy instrument for public health, is rather imperfect.\footnote{We reiterate that the main aim of the eviction moratoria was to keep people who lost their jobs because of the COVID-19 pandemic from also losing their homes.} Targeted policies such as mask wearing, social distancing and stay-at-home orders. are likely to be much more effective, as shown in \citeasnoun{pan2020covid}.  

\section{Background}

Understanding the economic, social, and health impacts of COVID-19, as well as the effects of various policies implemented to address the pandemic, is a crucial topic of research across multiple disciplines. However, a large scale multidisciplinary review of 102 articles attempting to estimate the impact of various COVID-19 policies on COVID-19 outcomes found that only \textit{one} of them met criteria and design checks for estimating causal impacts \cite{haber2021problems}. We therefore outline relevant background on the policy studied in this paper, eviction moratoria, to highlight the importance of carefully considering the methodological framework used for causal inference. 

\subsection{Eviction Moratoria and the Pandemic}

\begin{figure}
    \centering
    \caption{COVID-19 cases by moratorium status} 
    \includegraphics[width=0.8\textwidth]{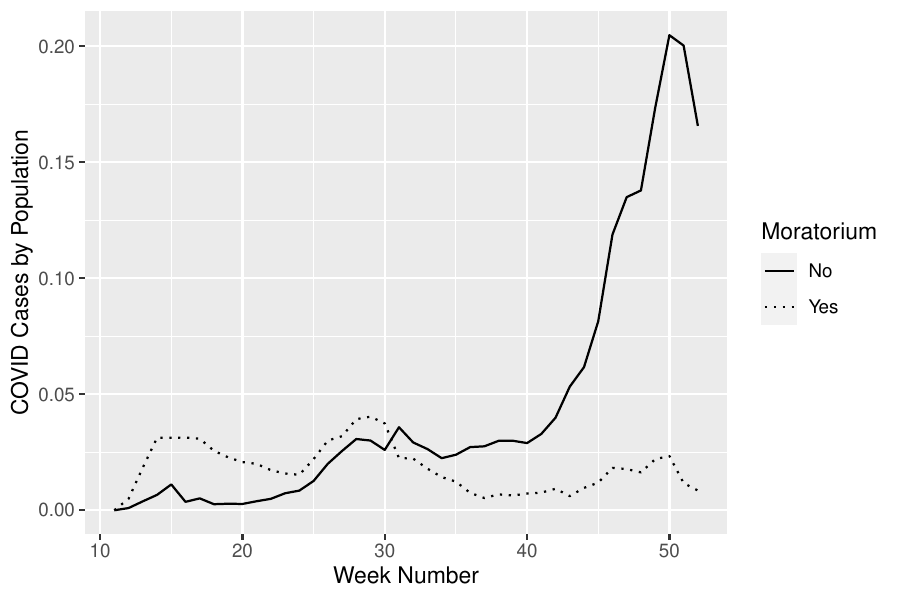}
    \caption*{This figure depicts the weekly ratio of new COVID-19 cases to total population, by county, for the year 2020. The dotted line represents the ratio for counties which had an eviction moratorium in place. The data, our main sample, is taken from the New York Times COVID-19 database, and consists of 59 counties from 30 US cities. See Section 3 for details.}
    \label{fig:moratoriatx}
\end{figure}

In the United States, one way that both federal and some state governments interceded to combat the spread of COVID-19 was by placing a moratorium on evictions. The justification for these moratoria was that evictions could lead to shelter overcrowding and homelessness as those forced to leave their homes searched for alternative housing. Thus, preventing landlords from evicting tenants would allow for better self-isolation, potentially limiting community spread. According to a CDC spokesperson,``it’s hard to follow social distancing orders if you have to double-up at a friend's or family member's house, and it's impossible if you're homeless and are forced to turn to shelters\footnote{Limited evidence indicates a wide degree of heterogeneity in the incidence of COVID-19 infections in homeless shelters during the initial weeks of the pandemic \cite{mosites2020assessment}.} as a last resort.''\footnote{https://www.vox.com/21569601/eviction-moratorium-cdc-covid-19-congress-rental-assistance-rent-crisis} Figure \ref{fig:moratoriatx} depicts the total number of per-county COVID-19 cases by population for our main sample, given the county's current weekly moratorium status. In total, there appears to be a much higher number of COVID-19 cases in counties without a current moratorium; however, this is not controlling for the crucially important presence of other COVID-19 mitigating policies. 

\begin{figure}
    \centering
    \caption{Average Eviction Filings by Week} 
    \includegraphics[width=0.7\textwidth]{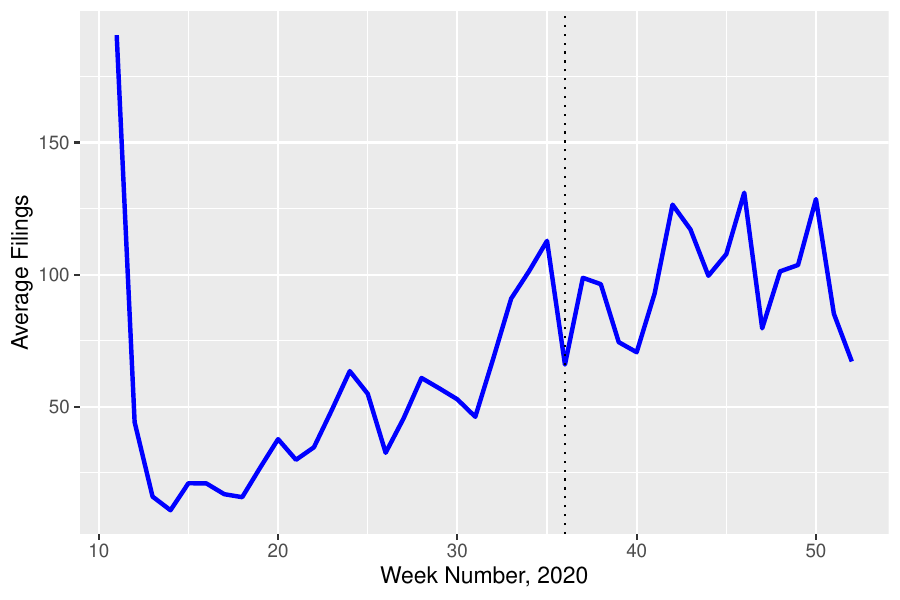}
    \caption*{This figure depicts the average number of eviction filings per week in 2020, for our sample of 30 US cities. The vertical dotted line shows the week of the CDC's national eviction moratorium going into effect. Data is taken from Princeton Eviction Lab \cite{evictionlab}.}
    \label{fig:filingsperweek}
\end{figure}

At the federal level, the CDC eviction moratorium went into effect on September 4th, 2020. %\footnote{Vox article: ``The CDC's eviction moratorium is an emergency measure; implemented after a partial eviction moratorium from the CARES Act expired at the end of July, the lack of congressional action left a void of federal protections against eviction. Some states and localities enacted renters protections, but fearing a national emergency, experts, advocates, and tenants began sounding the alarm."} 
Until January 1, 2021, landlords were no longer able to ``force tenants out of their homes due to a failure to pay rent, as long as the tenants \textbf{legally declare} they qualify for protection\footnote{In order to qualify for protection, tenants must have: used ``best efforts” to get ``all available” rent and housing assistance from the government, been below certain income thresholds, been unable to make rent because of a loss of household income, layoff, or ``extraordinary'' medical expenses, used ``best efforts'' to make partial rent payments, and demonstrated that eviction would make them homeless or force them to crowd into a new home.} under the order.'' %\footnote{See Vox article, footnote above} 
Landlords could still evict tenants for other reasons -- like ``engaging in criminal activity'' or ``threatening the health and safety of other residents.'' These requirements for obtaining protection under the national moratorium may explain why a substantial number of evictions still occurred even after September 4th. Alternatively, certain states or counties may have simply decided not to enforce the CDC ruling. Figure \ref{fig:filingsperweek} shows the average number of eviction filings in the Eviction Lab database by week in 2020 -- with no obvious effect of the September 4th ruling (depicted by the vertical line) for those counties in our sample. 

Since there is no formal indication as to whether or why certain counties decided to follow (or not follow) the national moratorium, like \citeasnoun{LEIFHEIT_ETAL:2021}, we perform our analysis at the local level instead of nationally. 

\subsection{Eviction Law in the United States}

In addition to heterogeneity in COVID response policies across state, there exists substantial heterogeneity in (pre-pandemic) state eviction statutes.\footnote{``Eviction Laws" Policy Surveillance Program of the LawAtlas Project} In most states, landlords must present tenants with written complaint (notice of intended eviction) for non-payment of rent a few days to a few weeks prior to the intended eviction date. Most, but not all, states then require court orders or judicial rulings in order for the physical eviction to proceed. If a tenant has the right to appeal the eviction, there is large variation across states in terms of the minimum number of days in which a trial can be scheduled after the tenant receives written notice. This means, in some states, landlords could have started eviction processes so that once moratoria lifted tenants could be removed expeditiously, and these removal processes differ according to underlying statutes.

In the context of COVID-related eviction moratoria, it is especially important to control for whether a state's laws require a landlord to waive the right to evict a tenant after accepting partial repayment of rent. Since part of the tenant's ``best efforts'' under the national moratorium require partial payment of rent if possible, states in which this prevents an eviction from going forward will have lower eviction rates (and potentially lower infection rates) as a result of the pre-existing eviction laws, not the COVID-related eviction policies. In addition, areas which had a moratorium on both eviction filings \textit{and} hearings saw more of a surge in evictions following expiration of local moratoria \cite{cowin2020measuring}. These examples of substantial variation make clear that any potential treatment and control groups for COVID-related policy are likely not comparable in terms of their underlying eviction policies - therefore we rely on ASC in our preferred analyses \cite{kreif2016examination}. We also conduct analysis on a subset of cities for which data on underlying eviction legislation is available in Section 6.  

\section{Data}

\begin{figure}[H]
    %\centering
    \caption{Total Counties with Moratoria Lifted per Week}
    \centering     %%% not \center
    \includegraphics[width=\textwidth]{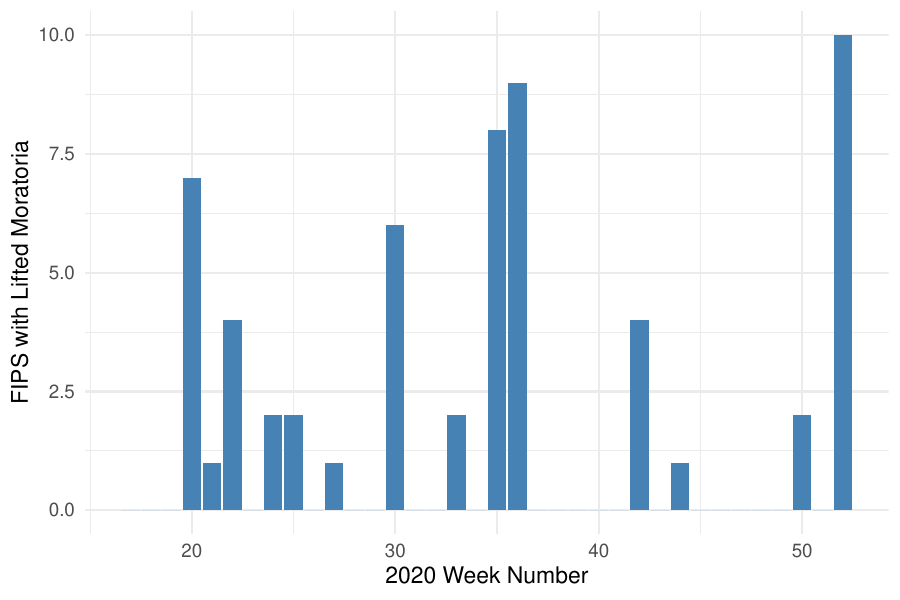}
    \caption*{This figure depicts the total number of counties in each week of the sample period which lifted moratoria in that week. The sample period begins April 20, 2020 and ends December 31, 2020, and the main sample consists of 59 counties from 30 US cities. Data on moratoria end dates is taken from Princeton Eviction Lab.}
    \label{fig:liftingsbyweek}
\end{figure}

Our sample contains 59 counties from the 30 US cities which enacted eviction moratoria and for which eviction data for 2020 is available on Princeton Eviction Lab. The sample period begins April 20, 2020 and ends December 31, 2020.\footnote{We begin the sample period in the first week in which all cities had active moratoria.} We extend the sample period to the end of 2020; even though the CDC eviction moratorium went into effect on September 4th, COVID-19 has a lag of 2-3 weeks, so we require data that goes past September to be able to properly extract cohort effects. We also do not have evidence that the nationwide moratorium made any difference at the local level on the actual number of evictions (see Figure \ref{fig:filingsperweek}). Since the Eviction Lab eviction data is at the city and/or county level, eviction moratorium information was also collected manually for each local municipality from this website. This is important to capture the true effect of moratorium endings, as there may be localities with orders that differ from their state's. For example, Texas's eviction moratorium ended on May 18, 2020, but the city of Austin, Texas, had an eviction moratorium in place through December 31st, 2020.  More concerning, some states may have had no state-level moratorium in place, yet certain metropolitan areas \textit{within} those states enacted their own orders. It is therefore crucial to collect detailed information about local municipality orders and not rely exclusively on state-level moratorium information. Since eviction data is at either the census-tract or the ZIP code level, all eviction counts were aggregated to the county level (using HUD USPS crosswalk information from Q1 2020). Figure \ref{fig:liftingsbyweek} depicts the number of counties in which moratoria lifted during each week of the sample period, and demonstrates no obvious pattern or grouping of the timing of moratoria endings across observations or with respect to the national CDC moratorium on September 4th, 2020. 

COVID-19 case and death information was taken from the New York Times database, which is provided at the county level in the {\tt covid19R} package available in the R statistical programming environment. Measurement errors in the data resulting in a few negative numbers for new cases and deaths were interpolated using a cubic spline. Demographic variables at the county level were taken from the 2018 American Community Survey, and include racial and ethnic demographics,\footnote{Which are known to be correlated with COVID-19 infection rates \cite{millett2020assessing,mahajan2020racial}, and are not controlled for by \citeasnoun{LEIFHEIT_ETAL:2021}.} educational attainment, average renting rates, and poverty and inequality indices. We also use Census estimates for population density in each county. OxCGRT provides a database of various COVID-19 policies at the state level, including start and end dates for mask mandates, stay-at-home orders, school closings, and an overall policy Stringency Index.\footnote{https://raw.githubusercontent.com/OxCGRT/USA-covid-policy/master/data/OxCGRT\_US\_latest.csv} County-level policy information was taken from the HHS.\footnote{healthdata.gov} Information on political party vote share was taken from the MIT Election Lab \cite{DVN/LYWX3D_2018}, and the Yale Climate Communication study \cite{yaleclimate} provides county-level survey data on belief in climate change, which we use as a proxy for trust in science. Finally, we merge selected details on eviction laws from the ``Eviction Laws'' Policy Surveillance Program of the LawAtlas Project to account for differences across eviction proceedings. 

Table \ref{tab:summarystats} presents summary statistics for selected variables; the high degree of variation in number of weekly eviction filings is of note. There is a strong negative correlation (-0.370) between local moratorium length and the number of eviction filings per county. We also note a weak negative correlation between the number of eviction filings and the strength of various other COVID-19 mitigating policies as captured in the local Stringency Index variable. The lack of correlation between moratorium length and political affiliation or stringency index is also of note. Also, the positive correlation between eviction filings and new COVID-19 cases is consistent with Figure \ref{fig:moratoriatx} (and the subsequent correlation with deaths).\footnote{Table \ref{tab:2} in Appendix \ref{app:b} contains a full correlation matrix for our policy and political variables.} 

\begin{table}[!htbp] \centering 
  \caption{Summary Statistics} 
  \label{tab:summarystats} 
\begin{tabular}{@{\extracolsep{5pt}}lcccc} 
\\[-1.8ex]\hline 
\hline \\[-1.8ex] 
Statistic & \multicolumn{1}{c}{Mean} & \multicolumn{1}{c}{St. Dev.} & \multicolumn{1}{c}{Min} & \multicolumn{1}{c}{Max} \\ 
\hline \\[-1.8ex] 
Moratorium Length (weeks) & 22.429 & 10.319 & 8 & 40 \\ 
Eviction Filings & 69.352 & 146.904 & 0 & 1,726 \\ 
Population Density & 2,087.582 & 2,541.545 & 99.106 & 13,801.320 \\ 
GINI Index & 0.479 & 0.035 & 0.383 & 0.562 \\ 
Percent White & 0.683 & 0.137 & 0.341 & 0.954 \\ 
Percent Black & 0.189 & 0.132 & 0.012 & 0.592 \\ 
Percent Latinx & 0.143 & 0.120 & 0.008 & 0.433 \\ 
Percent College Educated & 0.155 & 0.031 & 0.086 & 0.224 \\ 
Percent Renting  & 0.343 & 0.089 & 0.154 & 0.596 \\ 
Stringency Index & 52.567 & 12.677 & 8.330 & 79.630 \\ 
2016 Election Political Difference & 27.583 & 19.067 & 0.210 & 76.960 \\ 
Percent Belief in Climate Change & 70.437 & 5.640 & 57.360 & 81.238 \\ 
\hline \\[-1.8ex] 
\end{tabular} 
\caption*{\footnotesize This table presents summary statistics of various covariates for our sample of 30 US cities. Data on moratoria length and eviction filings is taken from Princeton Eviction Lab \cite{evictionlab}. Population density, GINI index, demographic and education data is taken from the 2018 ACS. COVID-19 stringency index is from the OxCGRT database. Political difference data is from the MIT Election Lab, and climate change belief data is from the Yale Climate Communication study.}
\end{table} 

\section{Methodology: Difference-in-Differences}

This section outlines the main econometric methods for staggered treatment timing settings used in this paper. We also present  negative binomial results following \citeasnoun{LEIFHEIT_ETAL:2021}, who did not account for cohort effects (as discussed earlier). 

For each method, our primary estimand of interest is the Average Treatment Effect on the Treated (ATT), $k$ periods after treatment:
\begin{equation}\label{eq:1}
    ATT_k \equiv \frac{1}{J} \sum_{j=1}^{J} Y_{j,T_j + k}(T_j) - Y_{j,T_j + k}(\infty).
\end{equation}
Event time relative to treatment time for unit $j$, $T_j$, is indexed by $k = t - T_j$. $Y_{j,T_j + k}(T_j)$ is the potential outcome at time $T_j + k$ under treatment, and Y$_{j,T_j + k}(\infty)$ is the potential outcome for untreated units. Their difference, $Y_{j,T_j + k}(T_j) - Y_{j,T_j + k}(\infty)$, is the individual (unit-level) treatment effect, which is averaged to obtain the $ATT$ as in Equation \eqref{eq:1}. 

\subsection{Negative Binomial Regression: Leifheit et al. (2021) Analysis}

For the state-level analysis, we follow \citeasnoun{LEIFHEIT_ETAL:2021} and use population-averaged negative binomial regression with two-way fixed effects (i.e. traditional difference-in-differences with an event study approach):
\begin{equation}
    Y_{it} = \alpha + \beta_1 \text{{T}}_{it} + \beta_2 \text{{Post}}_t + \beta_3 (\text{{T}}_{it} \times \text{{Post}}_t) + \gamma_i + \lambda_t + \epsilon_{it},
\end{equation}
with state-day as the unit of analysis, log of state population included as an offset, first-order autoregressive (AR1) structure, state and week fixed effects $\gamma_i$ and $\lambda_t$, and conventional (non-robust) standard errors. $\beta_3$ at various leads and lags from treatment time is the coefficient of interest for estimating $ATT_k$. 

\subsection{DR-DiD}

For our preferred DiD approach, we use the Double-Robust DiD (DR-DiD) proposed by \citeasnoun{callaway2021difference}. This semiparametric estimator corrects for the bias inherent in two-way fixed-effects event study estimates \cite{GOODMAN-BACON_MARCUS:2020}. 

The starting point for estimation in the DR-DiD model is the Group-Time ATT:
\begin{equation}
    ATT(g,t) = \mathbb{E}[Y_t(g) - Y_t(0)| G_g = 1],
\end{equation}
i.e., the ATT for units who are members of group $g$ at time period $t$. Nonparametric identification is obtained using the Double-Robust estimand of \citeasnoun{sant2020doubly}:
\begin{equation*}
    ATT(g,t;\delta) = \mathbb{E} \left[\left( 
    \frac{G_g}{\mathbb{E}[G_g]} - \frac{\frac{p_g(X)C}{1-p_g(X)}}{\mathbb{E}\left[\frac{p_g(X)C}{1-p_g(X)}\right]}
    \right) (Y_t - Y_{g-\delta-1} - m(X))
    \right] 
\end{equation*}
where $G_g = 1$ if a unit is first treated in period $g$, $C = 1$ if a unit is not treated in any time period (control), $p_g(X) = P(G_g = 1|X, G_g + C = 1)$ is the probability of being first treated in period $g$ conditional on covariates and either being a member of group $g$ or never treated, $m(X) = \mathbb{E}[Y_t - Y_{g-\delta-1} | X, C=1]$ is the outcome regression for the never-treated group, and $t = g - \delta - 1$ is the reference time period.\footnote{That is, the most recent time period when untreated potential outcomes are observed for group $g$.} This group-time ATT is then aggregated with respect to time-to-event $e$, using the weight of each cohort share and the associated influence function to obtain valid confidence intervals:
\begin{equation}\label{eq:agg}
    \theta_{es}(e) = \sum_{g \in G} {\textbf{1}} \{ g + e \leq T \} P(G = g | G + e \leq T) ATT(g,g+e).
\end{equation}

\subsection{Interaction-Weighted DID (IWES)}

Drawbacks of the DR-DiD procedure include the inability to include time-varying $X_i$, as all time-varying $X_i$ are held constant at their value in the last pre-treatment period. Further, in specifications with many controls the estimator does not converge due to propensity scores being very near 0 or 1.\footnote{This indicates the overlap condition may be violated, and alternatively, ASC may be more appropriate.} We therefore also perform an event study DiD using the \citeasnoun{SUN_ABRAHAM:2020} Interaction-Weighted estimator (IWES). This procedure is equivalent to the DR-DiD, except that the group-time ATT is estimated using a traditional two-way fixed effect regression \textbf{before} the weighted aggregation is performed. 

Specifically, the Group-Time ATTs $\beta_{g,e}$ are estimated:

\begin{equation}
    Y_{i,t} = \alpha + \sum_{g \in G} \sum_{g + e \neq -1} \beta_{g,e} ({\textbf{1}}(E_i = e) \cdot G^{g+e}_{i,t}) + \lambda_t + \epsilon_{it}
\end{equation}

using linear regression, and then aggregated as in Equation \eqref{eq:agg}. 

\subsection{Augmented Synthetic Control for Staggered Treatment Adoption}

The goal of synthetic control is to use the observed outcomes of $Y_{jt}$ to construct a weighted average of $Y_{iT}(\infty)$, which is not observed in our data. More specifically, synthetic control imputes the missing potential outcome as a weighted average of the control outcomes \cite{abadie2010synthetic,abadie2021using}. The weights are chosen as the solution to the constrained optimization problem:
\begin{align}\label{eq:SCM}
    \underset{\bm{\gamma}\in\Delta}{\min} ||\bm{V}^{1/2}\left(\bm{Y}_{i\cdot}-\tilde{\bm{ Y}}^\prime_{j\cdot}\right)||^2_2+\upsilon\sum\limits_{W_i=0}f(\gamma_i).\notag
\end{align}
where $\Delta$ is the appropriately sized simple. Synthetic control has many deep theoretical underpinnings, but at its core it is quite simple, to find a set of weights, $\bm\gamma$ that can be used to construct an estimator of the controls to be used as the appropriate counterfactual. In fact this simplicity in intuition is perhaps its greatest strength and one of the reasons for its perceived popularity. 

As \citeasnoun{abadie2010synthetic} show, when the treated units vector of lagged outcomes, $\bm Y_{i\cdot}$ lie interior of the convex hull of the control groups lagged outcomes $\tilde{\bm{ Y}}^\prime_{j\cdot}$ the corresponding weights will achieve perfect pre-treatment fit with the corresponding treatment effect estimator in possession of many desirable statistical properties. However, due to potential dimensionality issues, it is not universally feasible to achieve perfect pre-treatment fit. Even with close to perfect fit it is commonly recommended \cite{ABADIE_ETAL:2015} to run an extensive battery of placebo checks to ensure that $\bm\gamma$ do not overfit due to noise. ASC \citeaffixed{BEN-MICHAEL_FELLER_ROTHSTEIN:2021}{proposed by} adjusts for poor pre-treatment fit. 

\citeasnoun{BEN-MICHAEL_ETAL:2021} also extend SCM to the staggered treatment adoption setting. In this version, the original SCM estimator is considered for a single unit $j$. The SCM weights $\hat{\gamma_j}$ are the solution to:
\begin{equation}
    \underset{\bm{\gamma_{j}}\in\Delta_{j}^{scm}}{\min} \frac{1}{L_j} \left( \sum_{\ell=1}^{L_j} Y_{j,T_j -\ell} - \sum_{i=1}^{N}\gamma_{ij}Y_{i,T_j + \ell} \right )^2 + \lambda \sum_{n=1}^{N}\gamma_{ij}^2
\end{equation}
where $\gamma_{j}\in\Delta_{j}^{scm}$ has elements that satisfy $\gamma_{ij} \geq 0$ $\forall i$, $\sum_i \gamma_{ij} = 1$, and $\gamma_{ij} = 0$ whenever $i$ is not a possible donor. This modification focuses only on lagged outcomes and penalizes the weights towards uniformity using $\lambda$. 

Given the vector of weights $\hat{\gamma_{ij}}$ solving equation X, the estimate of the missing potential outcome for treated unit $j$ at event time $k$ is:
\begin{equation}
    \hat{Y}_{j,T_j + k}(\infty) = \sum_{i=1}^{N}\hat{\gamma_{ij}}Y_{j,T_j + k}
\end{equation}
and the estimated treatment effect is $\hat{\tau}_{jk} = Y_{j,T_j + k} - \hat{Y}_{j,T_j + k}(\infty)$, the difference between the observed outcome under treatment for the treated units and the estimated potential outcome for the synthetic control. 

With multiple treated units (i.e. the staggered adoption case), the above setup is generalized to create weights for each treated unit. The estimated treatment effect averages over the unit effect estimates:
\begin{equation}
    \hat{ATT_k} = \frac{1}{J}\sum_{j=1}^{J}\hat{\tau}_{jk}
\end{equation}
 which can be interpreted as both the average of individual unit SCM estimates, and an estimate for the average treated unit \cite{BEN-MICHAEL_FELLER_ROTHSTEIN:2021}. These equivalent interpretations are used to construct goodness-of-fit measures
 
 \begin{equation}
     q^{sep}(\hat{\Gamma}) \equiv \sqrt{\frac{1}{J} \sum_{j=1}^{J} \frac{1}{L_j} \sum_{\ell=1}^{L_j} \left(Y_{j,T_j -\ell} - \sum_{i=1}^{N}\gamma_{ij}Y_{i,T_j + \ell} \right )^2} 
 \end{equation}
and 
 \begin{equation}
     q^{pool}(\hat{\Gamma}) \equiv \sqrt{\frac{1}{L} \sum_{\ell=1}^{L} \left(\frac{1}{J} \sum_{T_j > \ell}Y_{j,T_j -\ell} - \sum_{i=1}^{N}\gamma_{ij}Y_{i,T_j + \ell} \right )^2}. 
 \end{equation}

The final ``partially pooled'' estimator minimizes a weighted average of these two measures:
\begin{equation}
    \nu (\hat{q}^{pool})^2 + (1-\nu)(\hat{q}^{sep})^2
\end{equation}
where $\hat{q}$ have been normalized by their values computed with weights $\hat{\Gamma}$. \citeasnoun{BEN-MICHAEL_ETAL:2021} describe a heuristic for choice of $\nu$ which we adhere to in our analysis.

\section{Results}

%\textbf{THIS WHOLE SECTION NEEDS TO BE REWRITTEN}
%Given the various covariates we have, we elected to investigate the impact of moratoria across a range of different specifications. To that end we consider 8 different specifications. Model 1 (C1 for cases, D1 for deaths) is our baseline specification and does not include any controls. Model 2 controls for (policy) stringency, mask mandates, and stay-at-home orders. Model 3 further includes a control for the number of weekly eviction filings while Model 4 adds the additional control of log population. Model 5 adds the percentage of individuals who are either black or renting. Model 6 includes the average number of eviction filings during the sample period instead of the number in the most recent week before treatment, as well as all controls from Model 5. Model 7 controls \textbf{only} for log population, average number of evictions, and political point difference. Model 8 adds further controls, depending on the method used. 

Given various cohort effects, we thought it easier to display our findings visually rather than in standard tabular form. For those interested, all specific point estimates and associated standard errors for both cases and deaths can be found in Appendix \ref{app:b} for all of the different estimation approaches deployed here. 

While we advocate for using the FIPS level, we first discuss state level results to help compare our findings with those of \citeasnoun{LEIFHEIT_ETAL:2021}. We use as covariates state-level COVID policies and the natural logarithm of the population. 

%Also, note that since Model 4 (C4/D4) nests both Models 2 and 3, we do not present the results from them graphically. This is to save space and that the main results were in no way impacted by progressing from C2 to C4 (D2 to D4). These specific figures are available upon request.

\subsection{State-Level Analysis}

\begin{figure}[h]
    %\centering
    %\label{fig:IWESmain}
    \caption{Effect of Moratorium End: State-level Analysis}
    \centering     %%% not \center
    \subfigure[Panel A: CASES]{\includegraphics[width=80mm]{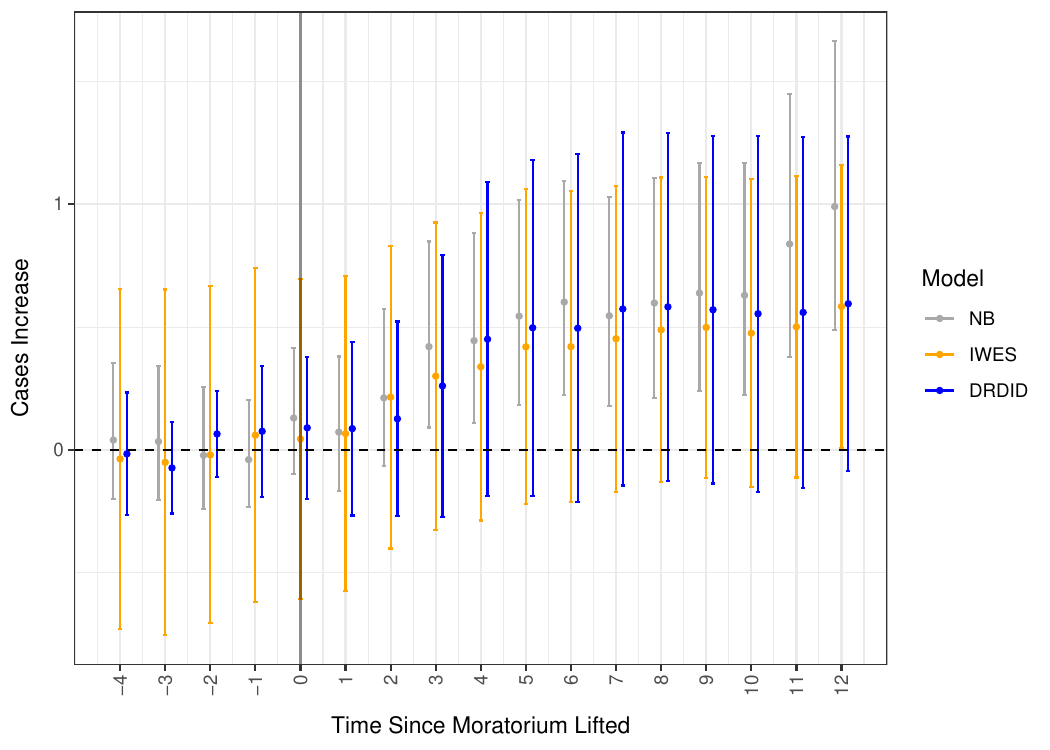}}
    \subfigure[Panel B: DEATHS]{\includegraphics[width=80mm]{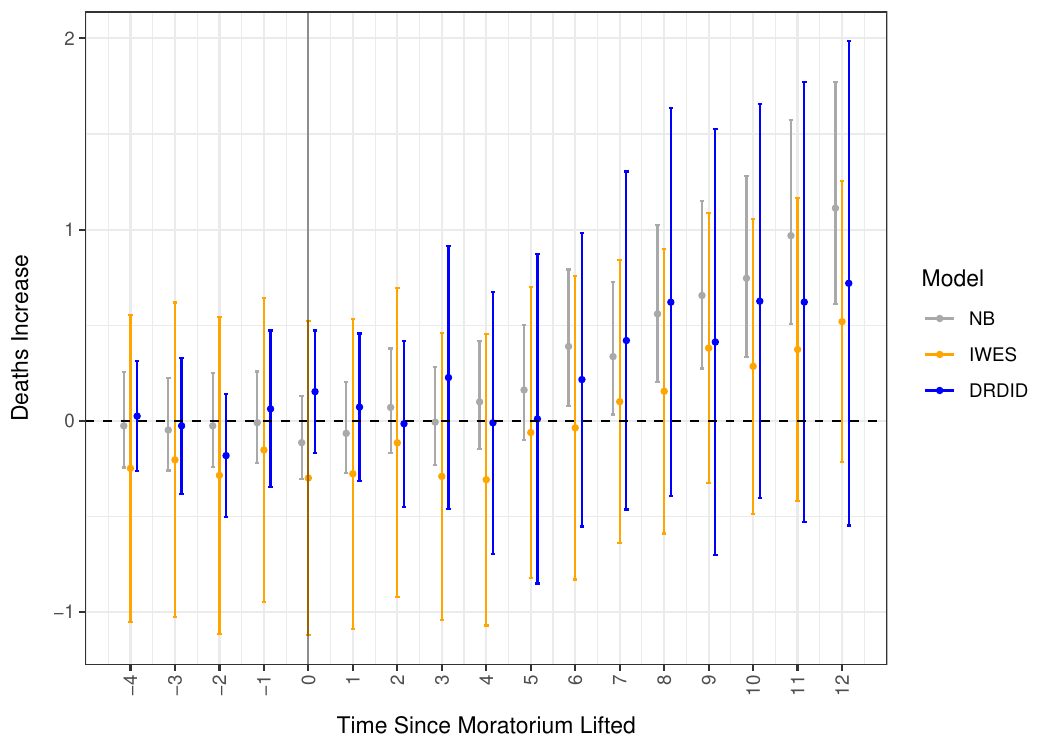}}
{\footnotesize \justifying \singlespacing{This figure compares results for three econometric methods at the state level. The model controls for state-level COVID-19 policies (stay-at-home orders, school closures, and mask mandates) and log population.} \par}
\label{fig:STATEmain}
\end{figure}

Figure \ref{fig:STATEmain} presents the cohort effects at the state level using the negative binomial specification of \citeasnoun{LEIFHEIT_ETAL:2021} as well as the IWES and the doubly robust DID estimators. For these estimators we include as controls state-level COVID-19 policies (measures of stay-at-home orders, school closures, and mask mandates) and the logarithm of the population. There are several striking and immediate features. After the expiration of a moratorium, cases go up. This is true for all three estimators. However, where they diverge is in the statistical strength of this increase. The Negative Binomial specification of \citeasnoun{LEIFHEIT_ETAL:2021} suggests statistically relevant increases in cases after 3 weeks. Neither the IWES or DR-DiD estimators find a statistically significant effect. Further, the estimated effects for the three estimators are quite similar after the lifting of the moratoria with the exception of weeks 11 and 12, where again the Negative Binomial estimator suggests another ``spike'' in cases. We view the near constancy of the impact of COVID-19 cases after about week 4 to be an equilibrium effect from the end of the moratorium being lifted. 

If we turn our attention to deaths, panel (b) in Figure \ref{fig:STATEmain} paints a much different figure at the state level. Initially, deaths at the state level fluctuate around zero until around week 6, when we start to see a dedicated increase. Again, the Negative Binomial specification finds statistically significant increases in deaths attributed to COVID-19 starting at week 6 whereas the IWES and DR-DiD estimators do not find statistically significant effects. The week 6 increase in deaths is intuitive given the roughly two week lag of COVID-19 effects. Thus, finding increases in COVID-19 cases after 3 weeks suggests that around week 5 or 6 an increases in deaths is expected. We also note that while there is more variation in deaths as we move further from the end of the moratorium being lifted, it does appear to be roughly stable, in line with the impact on cases. 

So, using the Negative Binomial specification promoted in \citeasnoun{LEIFHEIT_ETAL:2021} we see an increase in deaths from COVID-19 at around the same time (though our data run longer than their analysis) but we also find a more intuitive increase in cases. \citeasnoun{LEIFHEIT_ETAL:2021} claimed that their lack of finding an increase in cases prior to the spike in deaths was due to, among other things, undercounting of COVID cases (which could be attributed to those recently evicted not getting testing for COVID-19). 

\subsection{FIPS Level Analysis}

As argued earlier, the state level is not the appropriate level to address the impact of eviction moratoria on the spread and mortality of COVID-19 given the discrepancy between local and state ordinances. To that end we migrate from a state level aggregate dataset to a FIPS level analysis. Once we are in this setting we abandon the Negative Binomial specification advocated by \citeasnoun{LEIFHEIT_ETAL:2021} and focus our attention exclusively on the IWES and DR-DiD estimators.  

\begin{figure}[H]
    %\centering
    %\label{fig:IWESmain}
    \caption{Effect of Moratorium End, FIPS-level Benchmark}
    \centering     %%% not \center
    \subfigure[Panel A: CASES]{\includegraphics[width=80mm]{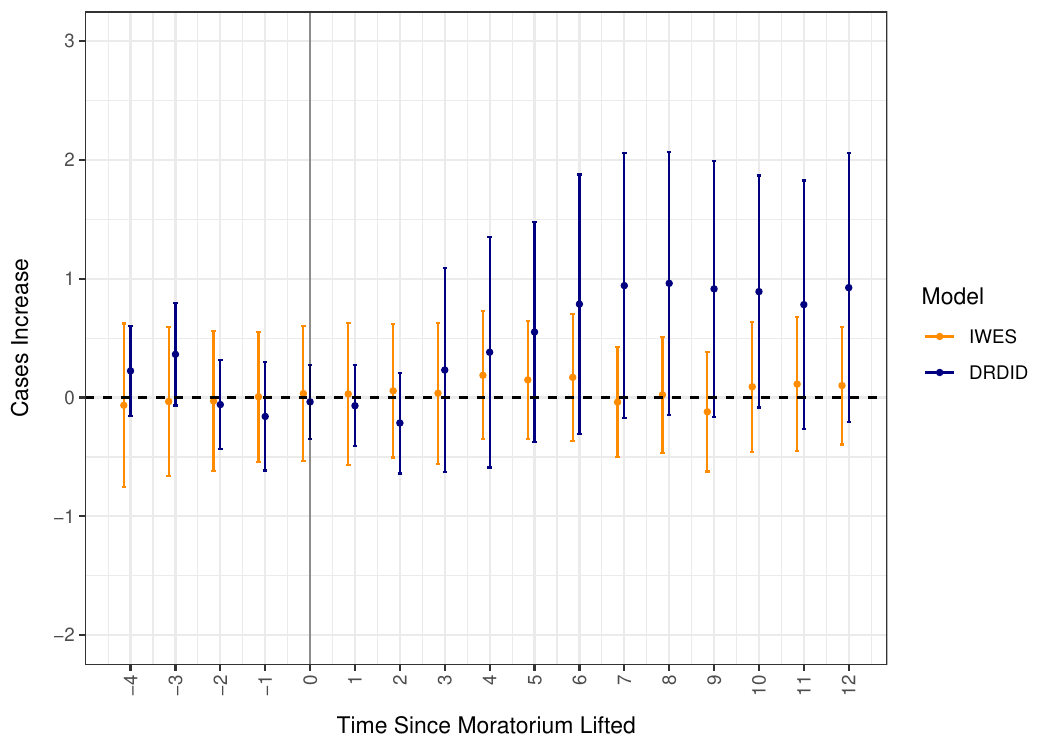}}
    \subfigure[Panel B: DEATHS]{\includegraphics[width=80mm]{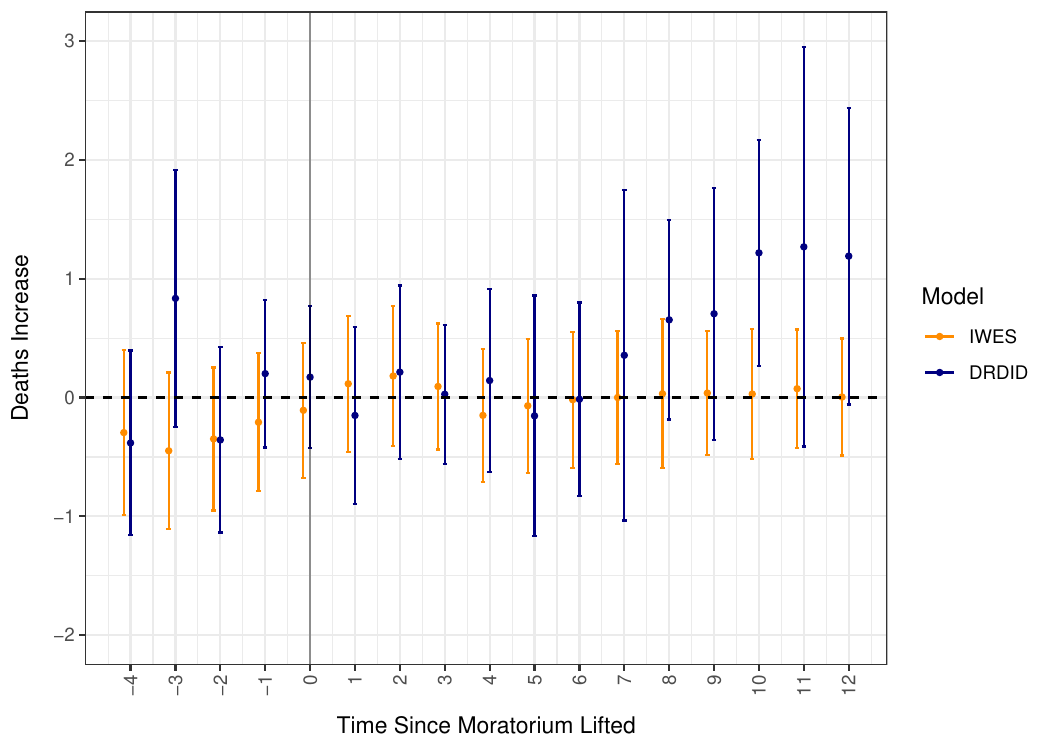}}
{\footnotesize \justifying \singlespacing{This figure depicts results using the Callaway \& Sant'Anna DR-DiD. \textbf{Panel A} shows estimates for cases. \textbf{Panel B} shows estimates for deaths. This benchmark model is identical across the IWES and DR-DiD specifications, and controls for average eviction filings, average stringency index, population, and demographic variables. Our outcome variables are the arcsine of the number of COVID-19 cases and deaths, therefore any values above zero are interpreted as an increase in the rate of new cases or deaths.} \par}
\label{fig:IWESmain}
\end{figure}

Our main specification for both estimators include the average stringency index (by FIPS), the logarithm of population, the proportion of the population that is black, the proportion of the population that is Hispanic, the proportion of the population that is college educated, and the average number of eviction filings (by FIPS). Figure \ref{fig:IWESmain} presents the cohort comparison of the benchmark specification across the IWES and the DR-DiD estimators. 

Several interesting features emerge. First, for cases, while both estimators fail to find statistically relevant effects due to the eviction moratoria expiring, the IWES estimator also finds a near 0 effect, while the DR-DiD estimator has a much larger positive effect which remains throughout the time-frame. We again see that in the first few weeks after the eviction moratoria expires at the the FIPS level, there are no noticeable impacts on cases, until around week 4, at which point the DR-DiD estimates experience the intuitive increase in average cases. Perhaps most interesting is that the simple switch from the state level to the FIPS level for the IWES estimates does not enjoy a similar increase in cases. Again this is additional evidence that buttresses our claim that the state level is the inappropriate focus for these effects. 

Turning our attention to deaths attributable to COVID-19, we see an expected pattern; the first few weeks after the eviction moratoria at the FIPS level is lifted there is no distinguishable pattern in deaths for either the IWES or DR-DiD estimates, and it is not until around week 8 that the DR-DiD estimates start to increase. We also see that even with the DR-DiD estimates increasing starting at week 8, aside from the significant effect at week 10, both estimates (IWES and DR-DiD) remain statistically insignificant throughout the time-frame, with the DR-DiD having economically larger (and positive) COVID-19 deaths. 

Having compared the estimated impacts of the eviction moratoria expiration at the FIPS level for a common specification to get a sense of the differences in the estimators, we now turn to different model specifications for each estimator. 

\subsubsection{Double-Robust DiD}

Focusing exclusively on the DR-DiD estimator, we consider three alternative specifications. Model 1 controls for the stringency index (held constant at the first pre-treatment period), a binary indicator for ever having mask mandate or stay-at-home orders, and the logarithm of the FIPS population. Model 2 is the same as Model 1 but includes the proportion of the population that is black, the proportion of the population that is Hispanic, the proportion of the population that is college educated. Model 3 is the benchmark model previously discussed (including average eviction filings in a FIPS to Model 2).  

\begin{figure}[h]
    %\centering
    %\label{fig:IWESmain}
    \caption{Effect of Moratorium End, FIPS-level}
    \centering     %%% not \center
    \subfigure[Panel A: CASES]{\includegraphics[width=80mm]{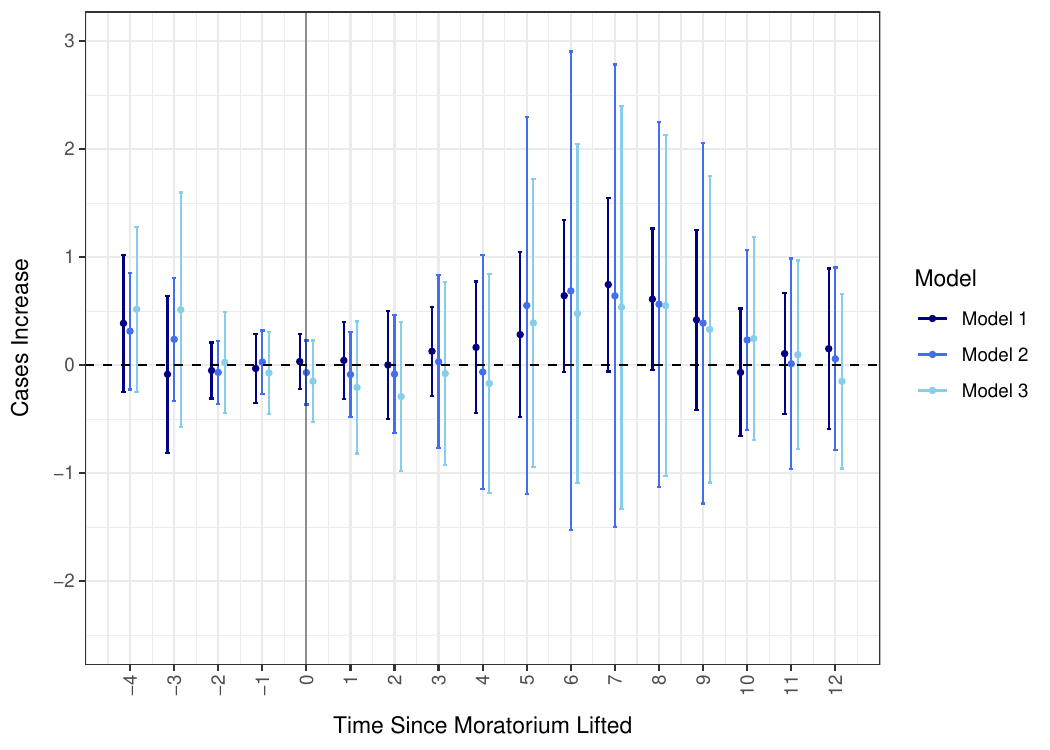}}
    \subfigure[Panel B: DEATHS]{\includegraphics[width=80mm]{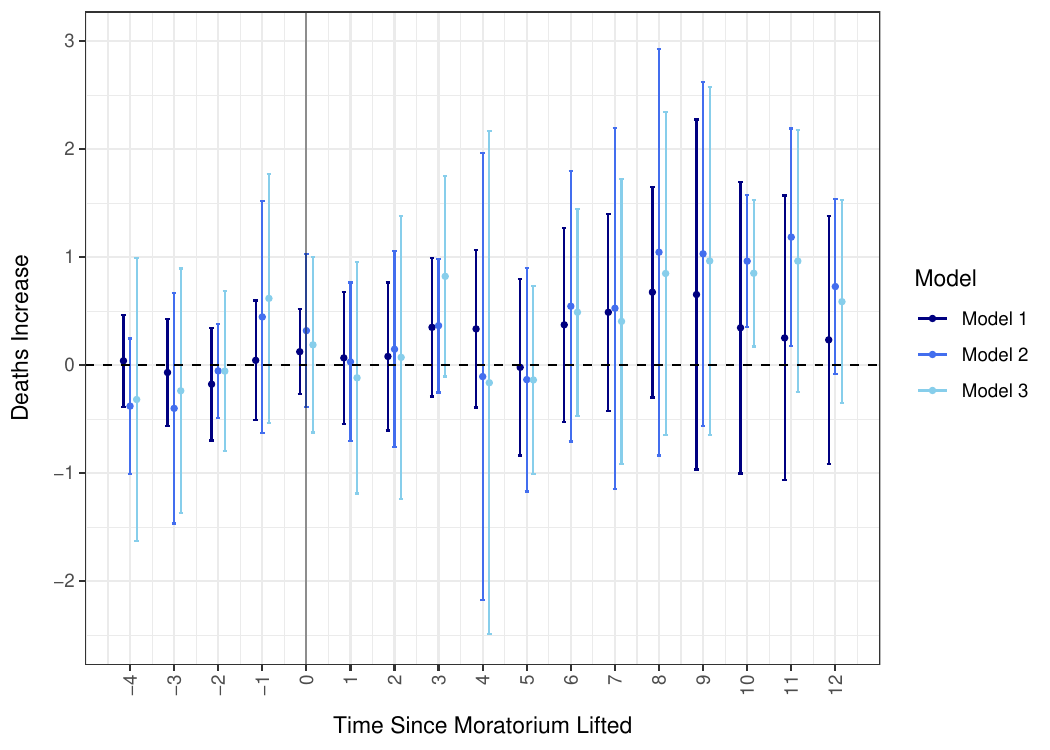}}
{\footnotesize \justifying \singlespacing{This figure depicts results using the Callaway \& Sant'Anna DR-DiD. \textbf{Panel A} shows estimates for cases. \textbf{Panel B} shows estimates for deaths. Model 1 includes policy and population controls. Model 2 adds demographic controls. Model 3 adds political controls and number of eviction filings. Our outcome variables are the arcsine of the number of COVID-19 cases and deaths, therefore any values above zero are interpreted as an increase in the rate of new cases or deaths. } \par}
\label{fig:DR-DiD_sub}
\end{figure}

Figure \ref{fig:DR-DiD_sub} presents the cohort effects across these three models. Several features are worth highlighting. All three specifications have similar patterns for cases, but with varying widths of confidence intervals around the point estimate, with the narrowest intervals stemming from Model 1. We also see a pronounced `bump' in cases starting around week 5 for all three specifications, peaking at week 7 and then slowly decaying back towards 0 by the end of the year. The increase in cases is intuitive but the lack of robust statistical findings is a bit concerning. All of the confidence intervals contain 0, speaking to the difficulty that is inherent in trying to discern the impact that the expiration of the eviction moratoria had on total COVID-19 cases. 

Turning our attention to deaths attributable to COVID-19, we see much less agreement across the three specifications. First, there is not a noticeable increase in the estimated increase in deaths until around week 8, but it is much less pronounced than it was for cases. We also see that for weeks 8 through 12, there is a difference between model specifications 2 and 3 and those from model 1 in terms of the magnitude of the number of deaths. Only in week 10 do we observe an estimated effect that is statistically different from 0, consistent with the totality of our findings so far.

\subsubsection{Interaction-Weighted DID}

\begin{figure}[H]
    %\centering
    %\label{fig:IWESmain}
    \caption{Effect of Moratorium End, FIPS-level: Sun \& Abraham IWES}
    \centering     %%% not \center
    \subfigure[Panel A: CASES]{\includegraphics[width=80mm]{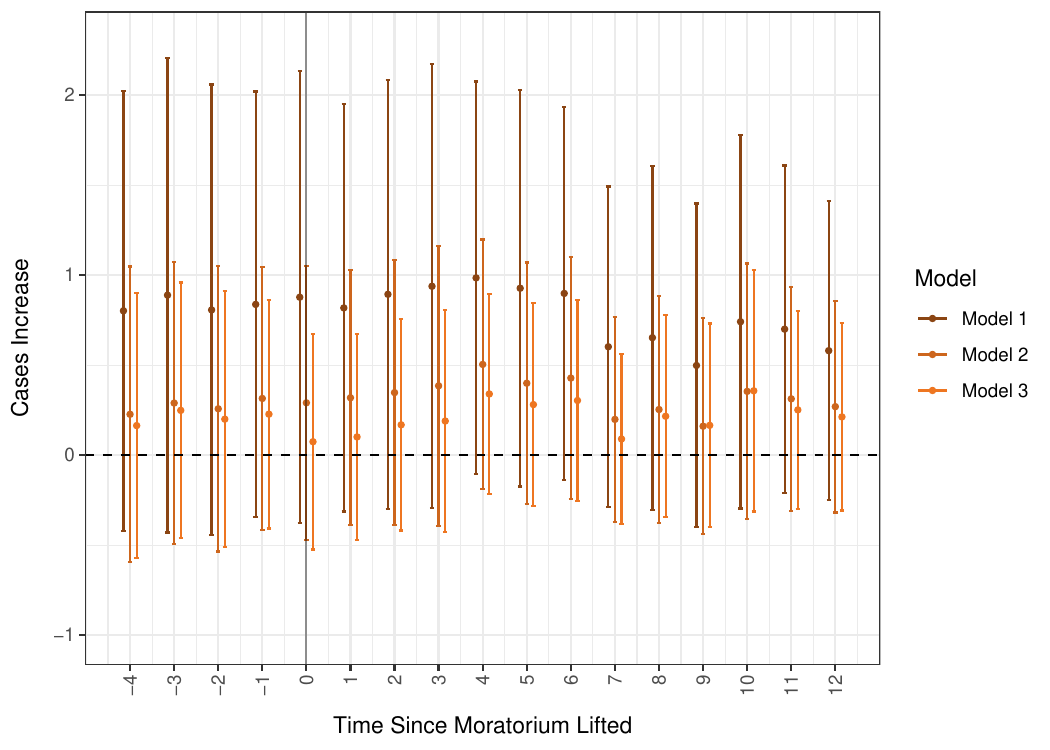}}
    \subfigure[Panel B: DEATHS]{\includegraphics[width=80mm]{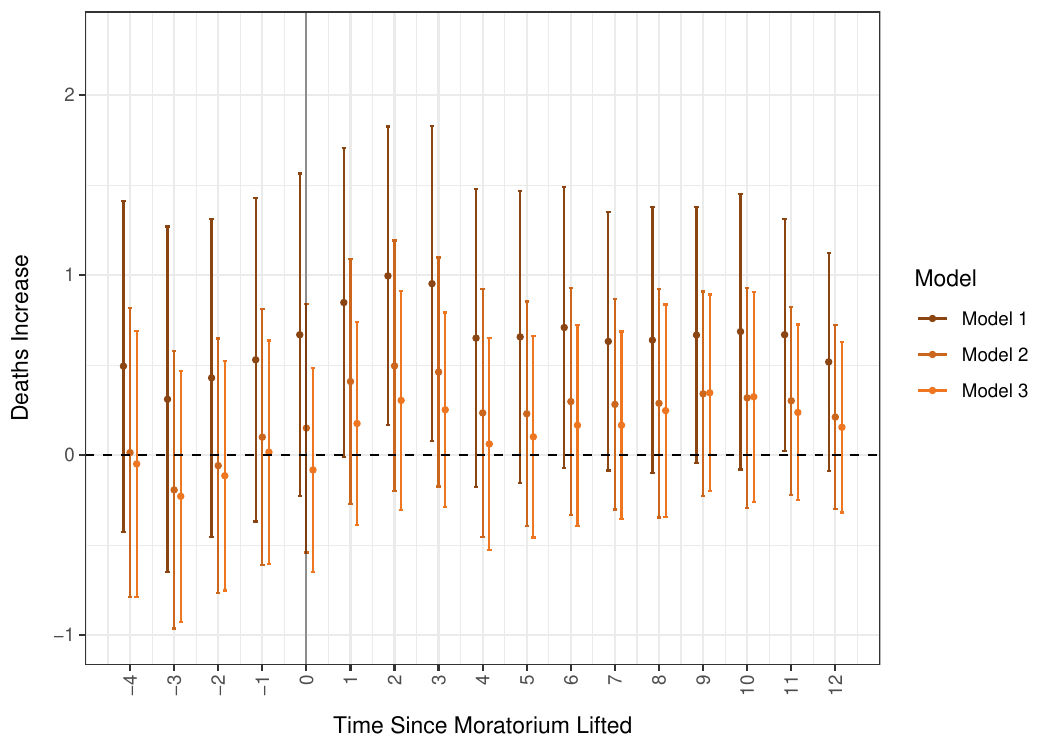}}
{\footnotesize \justifying \singlespacing{This figure depicts results using the Sun \& Abraham IWES. \textbf{Panel A} shows estimates for cases. \textbf{Panel B} shows estimates for deaths. Model 1 includes policy and population controls. Model 2 adds demographic controls. Model 3 adds political controls and number of eviction filings. Our outcome variable is the arcsine of the number of COVID-19 deaths, therefore any values above zero are interpreted as an increase in the rate of new deaths. } \par}
\label{fig:IWESsub}
\end{figure}

As the DR-DiD does not allow controlling for time-varying covariates, and often fails to converge when including the full suite of controls,\footnote{We note that the DR-DiD estimates show no statistical significance even when \textit{no controls} are used.} we now turn to the Interaction-Weighted DID (IWES). We report our findings in Figures \ref{fig:IWESsub} while the estimates and associated standard errors for cases and deaths can be found in Table \ref{table:IWESestimates} in Appendix \ref{app:b}. 

As with our analysis of the impacts using the IWES estimator, we consider three different specifications. We note that the specifications here are slightly different than those analyzed with the DR-DiD estimator since the estimators make slightly different assumptions on the nature of time variation in the controls. Here we consider a baseline model (Model 1) that includes as controls the stringency index (time-varying), county-level stay-at-home orders, county-level mask mandates, and the logarithm of the population in the FIPS. Model 2 includes all the controls from Model 1, but also includes the proportion of the population that is black, the proportion of the population that is Hispanic, the proportion of the population that is college educated. Finally, Model 3 adds to Model 2 with eviction filings (time varying) and political point difference in the 2016 election.

The results are consistent with our earlier DR-DiD estimates; there is no significant increase in COVID-19 cases (Panel A) following the lifting of eviction moratoria. Model 1 has the highest estimated impacts, which seem to occur immediately after the moratoria expire, but with wide confidence intervals containing 0. Models 2 and 3 display the same behavior, but with smaller estimated effects than Model 1 along with confidence intervals that contain 0. Interestingly, for all three model specifications we see that the estimated cohort effect drops at week 7. Overall the IWES estimator suggests that the expiration of the moratoria changed little regarding COVID-19 cases. 

Panel B of Figure \ref{fig:IWESsub} presents the results for deaths. Again, the results are nearly identical to the setup with cases. We see that the estimates from Model 1 are higher in magnitude than for Models 2 and 3, as to be expected, but the confidence intervals contain 0 (outside of weeks 1-3 for Model 1), throwing some doubt as to the true effect of these moratoria. We also can see higher estimated effects for the cohorts immediately after the moratoria are lifted, which is at odds with the behavior of the disease. If the eviction moratoria were implemented with the express intent of mitigating the spread of COVID, and the diseases as a one to two week lag time of transmission along with another one to two week lag time for severe symptoms to lead to death, then we would not anticipate such a large estimated effect for deaths so early on. This result also differs from the state level findings in \citeasnoun{LEIFHEIT_ETAL:2021}. 

Thus, across both the IWES and DR-DiD estimators for a variety of model specifications, we see increases in both cases of COVID and mortality from COVID, but with wide confidence intervals and time varying behavior that is not consistent with the behavior of the virus. This suggests that the identification of these eviction expiration effects are difficult to identify in practice, consistent with the concerns raised in \citeasnoun{GOODMAN-BACON_MARCUS:2020}.

\subsection{Robustness Checks}

Beyond our main DR-DiD and IWES specifications, we also consider if various forms of confounding and a different identification approach can more accurately reveal the impact of the expiration of evicition moratoria. To that end we consider a subset of our main dataset that incorporates observable features of eviction law at the FIPS/state level and the cohort effects using ASC with staggered adoption. 

\subsubsection{Augmented Synthetic Control with Staggered Adoption}

\begin{figure}[H]
    %\centering
    \caption{Effect of Moratorium End on New Cases and Deaths: Ben-Michael Augmented Synthetic Control \label{fig:ASC1}}
    \centering %%% not \center
    \subfigure[Cases]{\includegraphics[width=80mm]{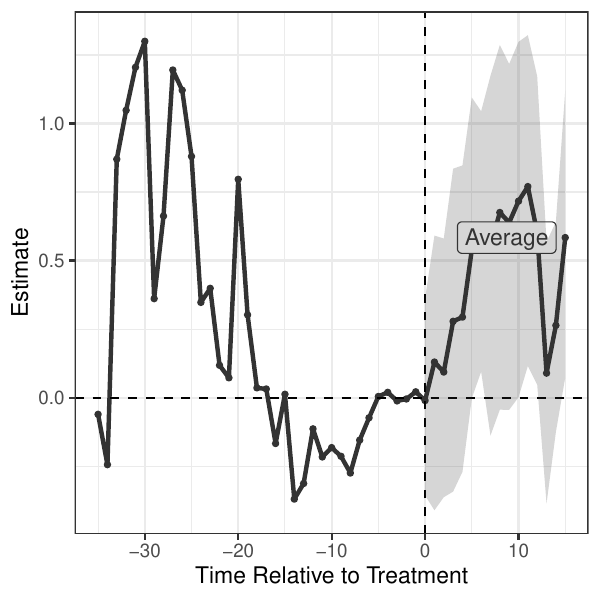}}
    \subfigure[Deaths]{\includegraphics[width=80mm]{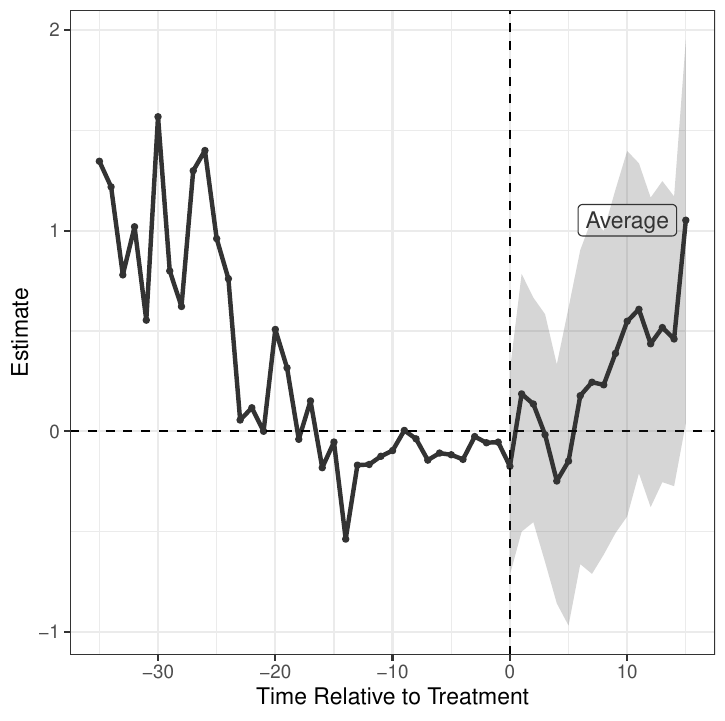}}
{\footnotesize \justifying \singlespacing{This figure depicts results using the Augmented Synthetic Control Method (Ben-Michael, 2020). The panels show the average effects by time relative to treatment for Cases (\textbf{Panel A}) and Deaths (\textbf{Panel B}). All moratoria are analyzed at the FIPS level. Our outcome variable is the arcsine of the number of COVID-19 cases/deaths, therefore any values above zero are interpreted as an increase in the rate of new cases/deaths.} \par}
\end{figure}

We now present the results for ASC with staggered adoption. Note that the current implementation in {\tt R} for the {\tt augsynth} package does not yet allow for matching on auxiliary covariates, so we report results without any matching. Figure \ref{fig:ASC1} depicts the results for total COVID cases and deaths attributable to COVID.\footnote{Figure \ref{fig:ASC2} in Appendix \ref{app:b} plots out the depicts pre-treatment balance and individual treatment effects for cases and deaths. Point estimates, standard errors, and confidence bounds are presented in Table \ref{tab:ASC}. } Panel A depicts average effects and demonstrates a very slight and temporarily significant increase in the incidence rate of cases six weeks after the moratoria lifted, and again in weeks 10-12 after lifting. 

Looking at Panel B, we again see no statistically meaningful effect of the moratoria expiration on deaths from COVID. We do see an increasing trend over time as we move further away from the moratoria ending, with a strange dip occurring at three weeks post expiration. The overall set of cohort effects is consistent with our earlier story that while there are estimated positive effects on mortality from COVID, said effects are difficult to precisely pin down. Our assumption is that if we were to also match on covariates that this would serve to further introduce noise as the quality of the mathces is likely to be poor as the number of covariates to match on increases. 

\subsubsection{Eviction Law Subset}

As there is a great deal of heterogeneity across both states and individual counties in terms of landlord and tenant protection statutes (see Section 2), we repeat the DR-DiD analysis here using a subset of data for which we have detailed information on existing tenancy laws. Arguably, it may be the case that areas with more tenant protection statutes in existence \textit{prior} to the COVID-19 pandemic may have been more likely to implement stricter eviction moratoria during the pandemic - introducing confounding. Including eviction law information reduces the size of our data to 17 cities and 36 counties. However, we are now able to control for whether a landlord waives the right to eviction if rent is partially repaid, the minimum number of days the landlord must provide before ending a tenancy due to non-payment of rent, the minimum number of days between when a landlord gives notice of tenancy termination and when the eviction may take place.\footnote{If there is time between notice and repossession, tenants may be able to file appeals or otherwise negotiate in order to avoid the eviction.} Once again, we find no evidence that the lifting of eviction moratoria increased the number of cases or deaths. 

\begin{figure}[H]
    %\centering
    \caption{Effect of Moratorium End on New Cases and Deaths: Eviction Law Subset}
    \centering     %%% not \center
    \subfigure[Cases]{\includegraphics[width=80mm]{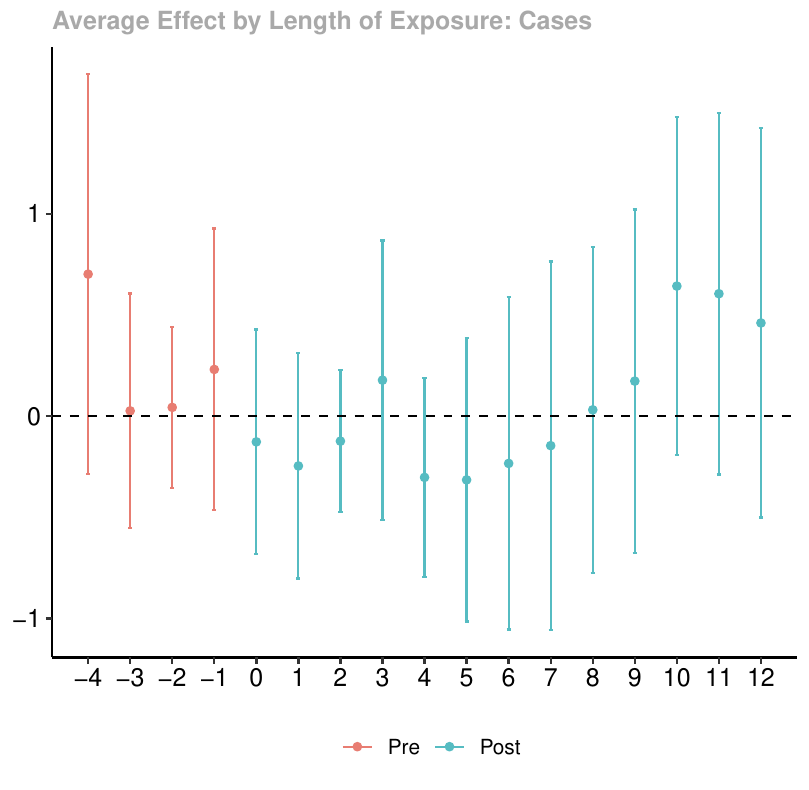}}
    \subfigure[Deaths]{\includegraphics[width=80mm]{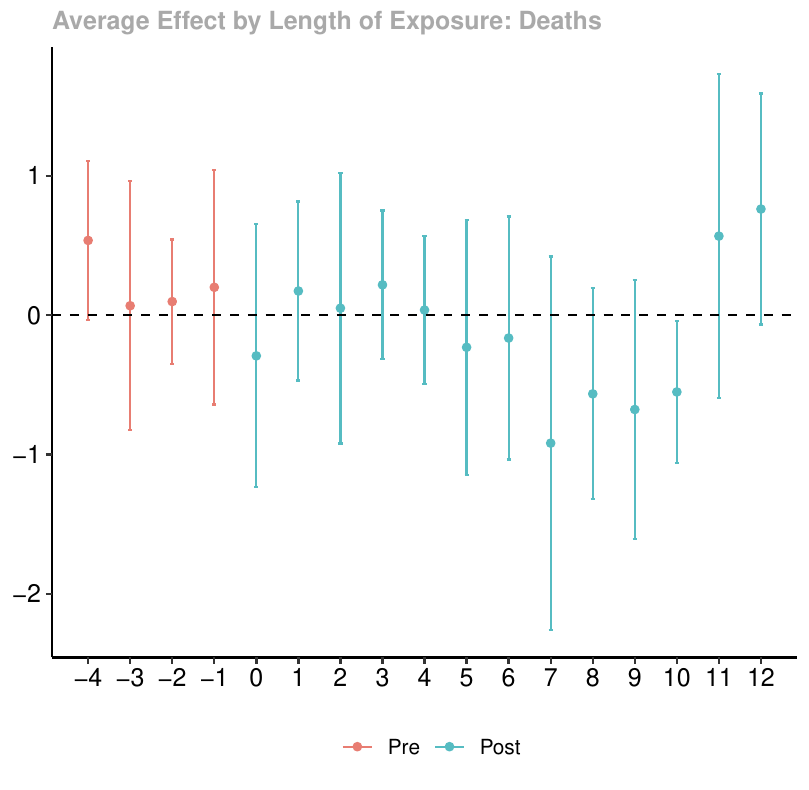}}
{\footnotesize \justifying \singlespacing{This figure depicts results for the subset of data for which we have information regarding eviction laws, using the Callaway Sant'Anna DR-DiD. The resulting sample contains 36 counties from 17 US cities. %\textbf{Panel A} shows estimates for cases, and \textbf{Panel B} for deaths. 
We control for whether a landlord waives the right to eviction if rent is partially repaid, the minimum number of days the landlord must provide before ending a tenancy due to non-payment of rent, the minimum number of days between when a landlord gives notice and when the eviction may take place, stringency index, and log of population. All moratoria are analyzed at the FIPS level.} \par}
\end{figure}

\section{Conclusions}

This paper set about critically examining the impact of local eviction moratoria on the spread of COVID-19. While several earlier studies documented increases in deaths attributable to COVID-19 following the expiration of these moratoria, we found minimal effects when deploying both newer econometric methods and what we believe to be more sensible data specification choices. Specifically, accounting for the differential timing of eviction moratoria across 44 states, switching from state level to country level data, controlling for number of evictions, and using cohort specific weighting for our time to treatment effects, we found that eviction moratoria likely did \textit{not} mitigate the spread of COVID-19. This finding was consistent across a range of specifications and estimation approaches. In fact, the only setting where we found an effect of the moratoria was when are data were aggregated up to the state level. 

However, as we stated earlier, the state is precisely the wrong level to focus attention on as different local municipalities had different rules in place for eviction filings and such aggregation washes away local variation in COVID-19 cases and deaths. Further, ignoring the fact that individuals could still be evicted when these eviction moratoria were in place represents a key omitted variable that helps to understand the impact of such a policy. Even the CDC's eviction moratoria, put in place on September 4th, 2020 nationally, did not prevent all evictions. Renters needed to qualify to seek eviction protections. 

These findings may seem to undermine the need for such moratoria, however, when they were initially instituted it was not as a non-pharmaceutical intervention per se, but a means to keep people from being homeless at a time of extreme economic uncertainty. Further, as the understanding of COVID-19 became more prevalent across the country, it is likely that individuals took other precautions to mitigate the risk of catching the virus and so eviction moratoria, kept in place as a means of reducing the spread of COVID-19, were simply not effective.  

Our crucial requirement of actual eviction numbers leads to the biggest limitation in this study: the sample size would ideally be larger than 30 cities. We believe, however, that the geographic dispersion of cities in the dataset is representative of the country as a whole. Future work may examine each city individually using a synthetic control method to uncover heterogeneity across cities or regions. In terms of methodology, we attempt to address the many issues inherent in policy impact evaluation for the COVID-19 pandemic by applying a variety of econometric techniques to our research question. However, the setting of staggered treatment timing is a fast-growing area of research, and it may be worthwhile for future authors to apply a staggered version of penalized synthetic control \cite{abadie2021penalized} or synthetic difference-in-differences \cite{arkhangelsky2021synthetic}, as they become available. 

We stress that our findings do not mean the moratoria were poor policies overall, far from it. The moratoria were designed not only to keep the spread of COVID-19 low, but to insulate individuals from losing their residence during this time of great upheaval. In that view the moratoria likely were quite effective. Indeed, \citeasnoun{an2021covid} find that the eviction moratoria reduced the financial stress of households by allowing them to redirect financial resources towards immediate consumption needs. Evictions in general lead to negative physical and mental health outcomes \cite**{desmond2015forced,benfer2021eviction}, a decreased likelihood of seeking medical attention \cite{collinson2018effects}, and damage to the overall public health of children \cite{schwartz2020cycles} - all of which are arguably even more problematic during a global pandemic. 

%\clearpage

\bibliographystyle{agsm}
\bibliography{COVID.bib}

\clearpage

\appendix 
\setcounter{table}{0}
\setcounter{figure}{0}
\renewcommand{\thetable}{A\arabic{table}}
\renewcommand{\thefigure}{A\arabic{figure}}

\section{ Leifheit Et Al. (2021) Replication}

For the initial replication at the state level, all information on moratoria and other policy start/end dates was collected as described in \citeasnoun{LEIFHEIT_ETAL:2021}. COVID-19 case and death incidence data by US state is from  the Covid Tracking Project.\footnote{covidtracking.com} Importantly, there are instances when the number of new cases and deaths is reported as a negative number, likely due to measurement error or data collection procedural changes. \citeasnoun{LEIFHEIT_ETAL:2021} do not describe these errors or how and whether they correct negative increases.\footnote{Some of these negative increases are on the order of thousands.} We interpolate all negative new cases and deaths using cubic interpolation. Population estimates for each state are taken from the 2018 American Community Survey. We create variables indicating $\geq$4, 3, 2 and 1 week prior, and 1, 2, 3, and $\geq$ 4 weeks after the start of mask mandates, school closures, and lifting of stay at home orders.\footnote{Arguably, for consistency it would be better to indicate the beginning of stay at home orders rather than their end, but we remain consistent with the Leifheit specification.} We create a variable of number of tests lagged by 7 days. Seven states which never imposed eviction moratoria are dropped from the analysis.\footnote{These states are Oklahoma, Arkansas, Georgia, Missouri, Ohio, South Dakota and Wyoming.} 

We do not code a state as having an active moratorium if the state does not pass measures with specific language regarding eviction proceedings. This differs from \citeasnoun{LEIFHEIT_ETAL:2021}, who code any state with civil court closures as having an eviction moratorium in place. In some states, courts are not necessarily involved for eviction proceedings to move forward. In order to perform additional analyses, we also extend the sample period to the end of 2020 \citeaffixed{LEIFHEIT_ETAL:2021}{from the original end date of September 3rd used in}. We add ACS estimates of racial and ethnic demographics, educational attainment, and poverty indicators at the state level. Finally, we add information on the percentage difference between Republican and Democrat vote share in the 2016 election.\footnote{ \tt github.com/kshaffer/election2016} Results for the negative binomial specification and the DR-DiD method are below.

\begin{table}[htbp]
  \centering
  \caption{Leifheit Replication (Negative Binomial): CASES}
  \adjustbox{max width=\textwidth}{%
    \begin{tabular}{cccccccccc}
    \hline\hline
          & Leifheit et. al &       & Cases &       &       &       &       &       &  \\
          &   (Appendix)    &       & (1)   & (2)   & (3)   & (4)   &       & (5)   & (6) \\
    Weeks post & IRR   &       & End Sept & End 2020 & By Week & All Covars &       & Log Incidence & Growth Rate \\
    \midrule
    1     & 0.93  &       & 1.201 & 1.130 & 1.065 & 1.017 &       & 0.0910 & -0.109 \\
          &       &       & (1.43) & (1.07) & (0.74) & (0.16) &       & (0.75) & (-0.48) \\
          &       &       &       &       &       &       &       &       &  \\
    2     & 0.99  &       & 1.168 & 1.073 & 1.067 & 0.918 &       & 0.112 & 0.162 \\
          &       &       & (1.06) & (0.55) & (0.55) & (-0.76) &       & (0.89) & (0.71) \\
          &       &       &       &       &       &       &       &       &  \\
    3     & 1.08  &       & 1.351 & 1.212 & 1.227 & 0.999 &       & 0.128 & 0.0817 \\
          &       &       & (1.96) & (1.45) & (1.44) & (-0.01) &       & (1.00) & (0.36) \\
          &       &       &       &       &       &       &       &       &  \\
    4     & 1.11  &       & 1.675** & 1.421** & 1.309 & 1.189 &       & 0.179 & 0.0408 \\
          &       &       & (3.28) & (2.62) & (1.66) & (1.49) &       & (1.40) & (0.18) \\
          &       &       &       &       &       &       &       &       &  \\
    5     & 1.16  &       & 1.658** & 1.445** & 1.350 & 1.208 &       & 0.143 & -0.175 \\
          &       &       & (3.14) & (2.72) & (1.67) & (1.62) &       & (1.11) & (-0.76) \\
          &       &       &       &       &       &       &       &       &  \\
    6     & 1.3   &       & 1.881*** & 1.545** & 1.485* & 1.281* &       & 0.248 & 0.222 \\
          &       &       & (3.86) & (3.19) & (2.02) & (2.11) &       & (1.92) & (0.96) \\
          &       &       &       &       &       &       &       &       &  \\
    7     & 1.34  &       & 2.092*** & 1.602*** & 1.508* & 1.315* &       & 0.252 & 0.925*** \\
          &       &       & (4.41) & (3.44) & (1.96) & (2.33) &       & (1.94) & (3.98) \\
          &       &       &       &       &       &       &       &       &  \\
    8     & 1.42  &       & 2.038*** & 1.546** & 1.563* & 1.256 &       & 0.248 & 0.206 \\
          &       &       & (4.16) & (3.15) & (2.02) & (1.91) &       & (1.89) & (0.88) \\
          &       &       &       &       &       &       &       &       &  \\
    9     & 1.47  &       & 2.073*** & 1.598*** & 1.605* & 1.289* &       & 0.233 & 0.0226 \\
          &       &       & (4.10) & (3.32) & (2.03) & (2.09) &       & (1.74) & (0.09) \\
          &       &       &       &       &       &       &       &       &  \\
    10    & \textbf{1.55} &       & 2.240*** & 1.639*** & 1.687* & 1.270 &       & 0.347** & -0.0992 \\
          &       &       & (4.33) & (3.46) & (2.14) & (1.94) &       & (2.58) & (-0.41) \\
          &       &       &       &       &       &       &       &       &  \\
    11    & \textbf{1.69} &       & 2.360*** & 1.630*** & 1.785* & 1.208 &       & 0.303* & -0.100 \\
          &       &       & (4.43) & (3.35) & (2.27) & (1.50) &       & (2.20) & (-0.41) \\
          &       &       &       &       &       &       &       &       &  \\
    12    & \textbf{1.92} &       & 2.936*** & 1.838*** & 1.926* & 1.323* &       & 0.385** & 0.127 \\
          &       &       & (5.32) & (4.15) & (2.47) & (2.20) &       & (2.79) & (0.52) \\
          \hline\hline
    \end{tabular}%
    }
  \label{tab:leifheitrepcases}%
\end{table}%

\begin{table}[htbp]
  \centering
  \caption{Leifheit Replication (Negative Binomial): DEATHS}
  \adjustbox{max width=\textwidth}{%
    \begin{tabular}{cccccccccc}
    \hline\hline
          & Leifheit et. al  &       & Deaths &       &       &       &       &       &  \\
          &   (Appendix)    &       & (1)   & (2)   & (3)   & (4)   &       & (5)   & (6) \\
    Weeks post & MRR   &       & End Sept & End 2020 & By Week & All Covars &       & Log Incidence & Growth Rate \\
    \midrule
    1     & 0.98  &       & 0.928 & 0.886 & 0.872 & 0.727** &       & 0.491 & 0.118 \\
          &       &       & (-0.53) & (-0.98) & (-1.04) & (-2.81) &       & (1.51) & (0.60) \\
          &       &       &       &       &       &       &       &       &  \\
    2     & 1.02  &       & 1.045 & 0.935 & 0.943 & 0.736** &       & 0.404 & 0.281 \\
          &       &       & (0.30) & (-0.52) & (-0.34) & (-2.58) &       & (1.22) & (1.43) \\
          &       &       &       &       &       &       &       &       &  \\
    3     & 1.12  &       & 1.320 & 1.070 & 1.051 & 0.816 &       & -0.199 & 0.294 \\
          &       &       & (1.82) & (0.53) & (0.25) & (-1.71) &       & (-0.58) & (1.46) \\
          &       &       &       &       &       &       &       &       &  \\
    4     & 1.16  &       & 1.184 & 0.993 & 0.981 & 0.832 &       & 0.587 & -0.0160 \\
          &       &       & (1.08) & (-0.05) & (-0.09) & (-1.54) &       & (1.71) & (-0.08) \\
          &       &       &       &       &       &       &       &       &  \\
    5     & 1.1   &       & 1.480* & 1.099 & 0.961 & 0.895 &       & 1.124** & -0.0283 \\
          &       &       & (2.50) & (0.73) & (-0.17) & (-0.94) &       & (3.22) & (-0.14) \\
          &       &       &       &       &       &       &       &       &  \\
    6     & 1.19  &       & 1.659** & 1.162 & 1.153 & 0.876 &       & 0.508 & 0.218 \\
          &       &       & (3.22) & (1.15) & (0.58) & (-1.11) &       & (1.44) & (1.04) \\
          &       &       &       &       &       &       &       &       &  \\
    7     & \textbf{1.64} &       & 2.086*** & 1.389* & 1.350 & 1.067 &       & 0.718* & 0.141 \\
          &       &       & (4.61) & (2.54) & (1.17) & (0.55) &       & (1.99) & (0.66) \\
          &       &       &       &       &       &       &       &       &  \\
    8     & \textbf{1.76} &       & 2.192*** & 1.336* & 1.346 & 1.011 &       & 0.347 & 0.571** \\
          &       &       & (4.85) & (2.21) & (1.13) & (0.09) &       & (0.95) & (2.63) \\
          &       &       &       &       &       &       &       &       &  \\
    9     & \textbf{1.98} &       & 2.790*** & 1.559*** & 1.510 & 1.096 &       & 0.675 & 0.313 \\
          &       &       & (6.15) & (3.34) & (1.52) & (0.75) &       & (1.75) & (1.37) \\
          &       &       &       &       &       &       &       &       &  \\
    10    & \textbf{2.73} &       & 3.023*** & 1.656*** & 1.694 & 1.154 &       & 0.179 & 0.0870 \\
          &       &       & (6.40) & (3.78) & (1.90) & (1.17) &       & (0.45) & (0.37) \\
          &       &       &       &       &       &       &       &       &  \\
    11    & \textbf{2.57} &       & 3.420*** & 1.746*** & 1.864* & 1.155 &       & 0.206 & 0.169 \\
          &       &       & (6.86) & (4.08) & (2.19) & (1.14) &       & (0.50) & (0.68) \\
          &       &       &       &       &       &       &       &       &  \\
    12    & \textbf{3.23} &       & 4.220*** & 1.969*** & 2.183** & 1.210 &       & 0.218 & 0.512* \\
          &       &       & (7.73) & (4.96) & (2.69) & (1.51) &       & (0.51) & (2.00) \\
          &       &       &       &       &       &       &       &       &  \\
    \hline\hline
    \end{tabular}%
    }
  \label{tab:addlabel}%
\end{table}%

\begin{table}
  \centering
  \caption{Leifheit Replication (DR Callaway Sant'Anna): CASES}
  \adjustbox{max width=\textwidth}{%
\begin{tabular}{l c c c c c c c}
\hline\hline
 & (1) & (2) & (3) & (4) & (5) & (6) & (7) \\
     Weeks Post & End Sept & End 2020 & Not-yet-Treated & All Covars(OR) & All Covars(DR)      & Log Incidence & Growth Rate \\
\hline
1  & $-0.00$          & $0.09$           & $0.10$           & $0.51$           & $2.90$           & $0.15$           & $-1.10$           \\
   & $ [-0.38; 0.37]$ & $ [-0.17; 0.34]$ & $ [-0.19; 0.40]$ & $ [-0.04; 1.07]$ & $ [-1.03; 6.83]$ & $ [-1.25; 1.55]$ & $ [ -4.44; 2.25]$ \\
   &       &       &       &       &       &       &       \\
2  & $-0.05$          & $0.13$           & $0.12$           & $0.53$           & $2.69$           & $0.06$           & $0.61$            \\
   & $ [-0.56; 0.46]$ & $ [-0.24; 0.50]$ & $ [-0.31; 0.55]$ & $ [-0.28; 1.33]$ & $ [-1.28; 6.66]$ & $ [-1.16; 1.28]$ & $ [ -0.58; 1.80]$ \\
   &       &       &       &       &       &       &       \\
3  & $0.02$           & $0.26$           & $0.21$           & $0.92$           & $2.61$           & $-0.29$          & $-2.26$           \\
   & $ [-0.49; 0.54]$ & $ [-0.28; 0.80]$ & $ [-0.18; 0.60]$ & $ [-0.29; 2.13]$ & $ [-3.54; 8.77]$ & $ [-2.18; 1.60]$ & $ [-10.38; 5.87]$ \\
   &       &       &       &       &       &       &       \\
4  & $0.15$           & $0.45$           & $0.30$           & $1.14$           & $2.37^{*}$       & $0.04$           & $-1.01$           \\
   & $ [-0.56; 0.86]$ & $ [-0.24; 1.14]$ & $ [-0.26; 0.86]$ & $ [-0.64; 2.91]$ & $ [ 0.48; 4.26]$ & $ [-1.79; 1.87]$ & $ [ -5.49; 3.46]$ \\
   &       &       &       &       &       &       &       \\
5  & $0.15$           & $0.50$           & $0.30$           & $1.32$           & $2.23$           & $0.69$           & $-0.77$           \\
   & $ [-0.86; 1.15]$ & $ [-0.19; 1.18]$ & $ [-0.33; 0.92]$ & $ [-0.27; 2.92]$ & $ [-0.96; 5.41]$ & $ [-1.22; 2.59]$ & $ [ -4.76; 3.21]$ \\
   &       &       &       &       &       &       &       \\
6  & $0.15$           & $0.50$           & $0.32$           & $1.41$           & $2.16$           & $0.77$           & $-1.15$           \\
   & $ [-0.62; 0.92]$ & $ [-0.19; 1.18]$ & $ [-0.40; 1.04]$ & $ [-0.34; 3.16]$ & $ [-1.45; 5.77]$ & $ [-0.95; 2.50]$ & $ [ -7.54; 5.25]$ \\
   &       &       &       &       &       &       &       \\
7  & $0.21$           & $0.57$           & $0.40$           & $1.40$           & $1.84$           & $1.00$           & $0.53$            \\
   & $ [-0.62; 1.03]$ & $ [-0.38; 1.53]$ & $ [-0.33; 1.13]$ & $ [-0.66; 3.46]$ & $ [-1.00; 4.68]$ & $ [-1.10; 3.10]$ & $ [ -0.49; 1.55]$ \\
   &       &       &       &       &       &       &       \\
8  & $0.15$           & $0.58$           & $0.40$           & $1.34$           & $1.81$           & $0.79$           & $0.52$            \\
   & $ [-0.79; 1.10]$ & $ [-0.06; 1.22]$ & $ [-0.45; 1.25]$ & $ [-0.85; 3.54]$ & $ [-1.12; 4.75]$ & $ [-2.18; 3.76]$ & $ [ -0.46; 1.49]$ \\
   &       &       &       &       &       &       &       \\
9  & $0.08$           & $0.57$           & $0.21$           & $1.92^{*}$       & $1.75$           & $0.69$           & $0.35$            \\
   & $ [-0.99; 1.15]$ & $ [-0.16; 1.30]$ & $ [-0.96; 1.38]$ & $ [ 0.35; 3.50]$ & $ [-2.33; 5.83]$ & $ [-1.29; 2.68]$ & $ [ -0.66; 1.36]$ \\
   &       &       &       &       &       &       &       \\
10 & $0.07$           & $0.55$           & $0.05$           & $1.93^{*}$       & $1.82$           & $0.91$           & $0.57$            \\
   & $ [-0.93; 1.06]$ & $ [-0.18; 1.29]$ & $ [-0.77; 0.88]$ & $ [ 0.22; 3.65]$ & $ [-0.87; 4.52]$ & $ [-1.42; 3.23]$ & $ [ -0.94; 2.08]$ \\
   &       &       &       &       &       &       &       \\
11 & $-0.06$          & $0.56$           & $0.04$           & $1.86$           & $2.01^{*}$       & $0.71$           & $-0.16$           \\
   & $ [-1.40; 1.28]$ & $ [-0.22; 1.34]$ & $ [-0.91; 0.99]$ & $ [-0.06; 3.79]$ & $ [ 0.77; 3.25]$ & $ [-1.70; 3.11]$ & $ [ -1.13; 0.82]$ \\
   &       &       &       &       &       &       &       \\
12 & $0.03$           & $0.60$           & $0.06$           & $2.12^{*}$       & $2.22^{*}$       & $0.85$           & $0.55$            \\
   & $ [-1.58; 1.63]$ & $ [-0.13; 1.32]$ & $ [-0.91; 1.04]$ & $ [ 0.66; 3.58]$ & $ [ 0.37; 4.08]$ & $ [-1.92; 3.62]$ & $ [ -0.47; 1.58]$ \\
\hline\hline
\end{tabular}
}
\label{table:coefficientsreplicationDRDID}
\end{table}

\begin{table}
  \centering
  \caption{Leifheit Replication (DR Callaway Sant'Anna): DEATHS}
  \adjustbox{max width=\textwidth}{%
\begin{tabular}{l c c c c c c c}
\hline\hline
 & (1) & (2) & (3) & (4) & (5) & (6) & (7) \\
     Weeks Post & End Sept & End 2020 & Not-yet-Treated & All Covars(OR) & All Covars(DR)      & Log Incidence & Growth Rate \\
    \midrule
1  & $0.03$           & $0.07$           & $0.09$           & $0.41$           & $1.83$           & $-0.16$          & $-0.23$          \\
   & $ [-0.32; 0.38]$ & $ [-0.35; 0.49]$ & $ [-0.26; 0.45]$ & $ [-0.36; 1.18]$ & $ [-0.28; 3.95]$ & $ [-1.53; 1.22]$ & $ [-1.09; 0.63]$ \\
   &       &       &       &       &       &       &       \\
2  & $-0.10$          & $-0.01$          & $0.10$           & $0.53$           & $2.68^{*}$       & $0.02$           & $0.38$           \\
   & $ [-0.63; 0.43]$ & $ [-0.45; 0.42]$ & $ [-0.29; 0.50]$ & $ [-0.58; 1.65]$ & $ [ 0.30; 5.06]$ & $ [-1.53; 1.56]$ & $ [-1.14; 1.90]$ \\
   &       &       &       &       &       &       &       \\
3  & $0.27$           & $0.23$           & $0.37$           & $0.70$           & $2.78^{*}$       & $0.38$           & $-0.16$          \\
   & $ [-0.62; 1.15]$ & $ [-0.43; 0.89]$ & $ [-0.33; 1.07]$ & $ [-0.47; 1.86]$ & $ [ 1.38; 4.18]$ & $ [-1.42; 2.17]$ & $ [-1.02; 0.70]$ \\
   &       &       &       &       &       &       &       \\
4  & $-0.01$          & $-0.01$          & $0.08$           & $0.84$           & $3.27$           & $0.14$           & $-0.27$          \\
   & $ [-0.69; 0.66]$ & $ [-0.64; 0.62]$ & $ [-0.53; 0.68]$ & $ [-0.44; 2.12]$ & $ [-0.17; 6.70]$ & $ [-1.36; 1.63]$ & $ [-1.16; 0.62]$ \\
   &       &       &       &       &       &       &       \\
5  & $-0.06$          & $0.01$           & $0.01$           & $1.13$           & $3.06$           & $0.37$           & $-0.12$          \\
   & $ [-0.79; 0.67]$ & $ [-0.85; 0.87]$ & $ [-0.69; 0.71]$ & $ [-0.87; 3.13]$ & $ [-0.98; 7.10]$ & $ [-1.09; 1.83]$ & $ [-0.91; 0.68]$ \\
   &       &       &       &       &       &       &       \\
6  & $0.08$           & $0.22$           & $0.13$           & $1.12$           & $2.83$           & $1.42$           & $0.41$           \\
   & $ [-0.94; 1.10]$ & $ [-0.55; 0.99]$ & $ [-0.55; 0.81]$ & $ [-0.81; 3.06]$ & $ [-1.13; 6.79]$ & $ [-0.01; 2.86]$ & $ [-1.15; 1.97]$ \\
   &       &       &       &       &       &       &       \\
7  & $0.16$           & $0.42$           & $0.22$           & $1.53$           & $3.03$           & $1.25$           & $-0.26$          \\
   & $ [-0.83; 1.16]$ & $ [-0.34; 1.18]$ & $ [-0.50; 0.93]$ & $ [-0.71; 3.78]$ & $ [-1.26; 7.32]$ & $ [-0.35; 2.85]$ & $ [-1.13; 0.61]$ \\
   &       &       &       &       &       &       &       \\
8  & $0.56$           & $0.62$           & $0.50$           & $1.20$           & $2.84$           & $1.73^{*}$       & $0.56$           \\
   & $ [-0.66; 1.78]$ & $ [-0.47; 1.71]$ & $ [-0.39; 1.39]$ & $ [-1.39; 3.80]$ & $ [-0.93; 6.61]$ & $ [ 0.20; 3.27]$ & $ [-1.43; 2.56]$ \\
   &       &       &       &       &       &       &       \\
9  & $0.17$           & $0.41$           & $0.23$           & $2.27$           & $2.87$           & $1.14$           & $-0.36$          \\
   & $ [-1.12; 1.45]$ & $ [-0.75; 1.57]$ & $ [-0.67; 1.13]$ & $ [-0.48; 5.02]$ & $ [-1.04; 6.78]$ & $ [-0.46; 2.74]$ & $ [-1.23; 0.52]$ \\
   &       &       &       &       &       &       &       \\
10 & $0.37$           & $0.63$           & $0.32$           & $2.37$           & $2.56$           & $1.38$           & $-0.12$          \\
   & $ [-0.87; 1.61]$ & $ [-0.48; 1.73]$ & $ [-0.70; 1.35]$ & $ [-0.33; 5.08]$ & $ [-2.14; 7.26]$ & $ [-0.49; 3.24]$ & $ [-0.93; 0.69]$ \\
   &       &       &       &       &       &       &       \\
11 & $0.43$           & $0.62$           & $0.35$           & $2.71^{*}$       & $2.39$           & $1.60$           & $-0.29$          \\
   & $ [-0.72; 1.58]$ & $ [-0.53; 1.78]$ & $ [-0.73; 1.43]$ & $ [ 0.22; 5.21]$ & $ [-1.10; 5.88]$ & $ [-0.97; 4.16]$ & $ [-1.39; 0.81]$ \\
   &       &       &       &       &       &       &       \\
12 & $0.43$           & $0.72$           & $0.35$           & $2.74$           & $2.04$           & $1.95^{*}$       & $0.23$           \\
   & $ [-0.83; 1.69]$ & $ [-0.44; 1.88]$ & $ [-0.89; 1.58]$ & $ [-0.50; 5.99]$ & $ [-1.75; 5.84]$ & $ [ 0.09; 3.82]$ & $ [-0.90; 1.36]$ \\
\hline\hline
\end{tabular}
}
\label{table:coefficients}
\end{table}

\section{Tables and Figures \label{app:b}}

Table \ref{tab:2} examines the correlation between the length of a county's eviction moratorium and other variables reported on in the text. 

{\begin{table}
\caption{Correlations: Policy and Political Variables}
\label{tab:2}
\begin{adjustbox}{max width=\textwidth}
\def\sym#1{\ifmmode^{#1}\else\(^{#1}\)\fi}
\begin{tabular}{l*{6}{c}}
\hline\hline                                                                                          \\
          &Eviction Filings         & Stringency Index         &Political Difference         &New Cases         &New Deaths         &   Moratorium Length         \\
\hline
Eviction Filings&        1         &                  &                  &                  &                  &                  \\
Stringency Index&   -0.289\sym{*}  &        1         &                  &                  &                  &                  \\
Political Difference&   -0.289\sym{*}  &   0.0508         &        1         &                  &                  &                  \\
New Cases &    0.895\sym{***}&  -0.0784         &   -0.282\sym{*}  &        1         &                  &                  \\
New Deaths&    0.738\sym{***}&   0.0778         &   -0.214         &    0.919\sym{***}&        1         &                  \\
Moratorium Length    &   -0.370\sym{**} &   0.0670         &    0.157         &   -0.235         &   -0.105         &        1         \\
\hline\hline
\multicolumn{7}{l}{\footnotesize \sym{*} \(p<0.05\), \sym{**} \(p<0.01\), \sym{***} \(p<0.001\)}\\
\end{tabular}
\end{adjustbox}
\end{table}
}

\begin{table}[ht]
\centering
\caption{Summary Statistics by City}
\adjustbox{max width=\textwidth}{%
\begin{tabular}{rlrrrrrrrrr}
  \hline\hline
 &  & Eviction & Moratorium & Stringency & \%  White & \% Black & \% Hispanic & \% College & \% Rented & Political \\ 
  & City & Filings & Length & Index & & & & & & Point Diff \\ 
  \hline
1 & Austin, TX & 2.33 & 40.00 & 52.84 &  &  &  &  &  & 30.99 \\ 
  2 & Boston, MA & 5.02 & 26.00 & 59.92 & 0.69 & 0.12 & 0.12 & 0.19 & 0.42 & 43.02 \\ 
  3 & Bridgeport, CT & 12.40 & 35.00 & 61.80 & 0.71 & 0.11 & 0.20 & 0.18 & 0.32 & 20.29 \\ 
  4 & Charleston, SC & 91.00 & 9.00 & 45.28 & 0.69 & 0.26 & 0.05 & 0.21 & 0.32 & 7.86 \\ 
  5 & Cincinnati, OH & 130.07 & 11.00 & 59.26 & 0.67 & 0.25 & 0.03 & 0.16 & 0.38 & 9.52 \\ 
  6 & Cleveland, OH & 68.93 & 13.00 & 58.91 & 0.63 & 0.29 & 0.06 & 0.14 & 0.37 & 34.99 \\ 
  7 & Columbus, OH & 207.48 & 11.00 & 58.91 & 0.65 & 0.23 & 0.06 & 0.17 & 0.43 & 25.89 \\ 
  8 & Dallas, TX & 276.60 & 8.00 & 52.13 & 0.61 & 0.23 & 0.41 & 0.13 & 0.47 & 26.24 \\ 
  9 & Fort Worth, TX & 120.65 & 8.00 & 52.13 & 0.70 & 0.13 & 0.24 & 0.17 & 0.35 & 14.44 \\ 
  10 & Gainesville, FL & 15.19 & 16.00 & 49.60 & 0.69 & 0.20 & 0.10 & 0.15 & 0.33 & 22.52 \\ 
  11 & Greenville, SC & 139.62 & 9.00 & 45.28 & 0.72 & 0.18 & 0.09 & 0.16 & 0.29 & 24.70 \\ 
  12 & Hartford, CT & 17.98 & 35.00 & 61.80 & 0.71 & 0.14 & 0.18 & 0.15 & 0.32 & 21.51 \\ 
  13 & Houston, TX & 239.21 & 8.00 & 52.13 & 0.70 & 0.16 & 0.34 & 0.14 & 0.34 & 17.52 \\ 
  14 & Indianapolis, IN & 209.05 & 21.00 & 47.78 & 0.60 & 0.29 & 0.11 & 0.14 & 0.41 & 22.83 \\ 
  15 & Jacksonville, FL & 122.88 & 18.00 & 49.60 & 0.60 & 0.29 & 0.10 & 0.13 & 0.40 & 1.50 \\ 
  16 & Kansas City, MO & 67.00 & 10.00 & 45.09 & 0.68 & 0.23 & 0.09 & 0.14 & 0.37 & 16.64 \\ 
  17 & Las Vegas, NV & 323.31 & 29.00 & 47.66 & 0.58 & 0.12 & 0.31 & 0.11 & 0.41 & 10.65 \\ 
  18 & Memphis, TN & 258.17 & 11.00 & 48.29 & 0.39 & 0.54 & 0.07 & 0.12 & 0.39 & 27.63 \\ 
  19 & Milwaukee, WI & 147.10 & 8.00 & 44.59 & 0.58 & 0.26 & 0.15 & 0.13 & 0.47 & 37.45 \\ 
  20 & Minn.-Saint Paul, MN & 5.11 & 39.00 & 55.73 & 0.69 & 0.13 & 0.07 & 0.20 & 0.37 & 37.38 \\ 
  21 & New Orleans, LA & 34.98 & 13.00 & 49.93 & 0.34 & 0.59 & 0.06 & 0.15 & 0.42 & 66.16 \\ 
  22 & New York, NY & 0.11 & 13.00 & 72.13 & 0.65 & 0.13 & 0.21 & 0.17 & 0.28 & 19.04 \\ 
  23 & Philadelphia, PA & 71.10 & 23.00 & 52.20 & 0.39 & 0.42 & 0.15 & 0.12 & 0.42 & 66.91 \\ 
  24 & Phoenix, AZ & 538.48 & 32.00 & 42.00 & 0.79 & 0.06 & 0.31 & 0.14 & 0.34 & 3.45 \\ 
  25 & Pittsburgh, PA & 7.97 & 23.00 & 52.64 &  &  & 0.01 & 0.13 & 0.24 & 30.37 \\ 
  26 & Richmond, VA & 13.24 & 24.00 & 49.77 &  &  &  &  &  & 33.39 \\ 
  27 & South Bend, IN & 18.36 & 21.00 & 47.78 & 0.77 & 0.13 & 0.09 & 0.12 & 0.30 & 0.21 \\ 
  28 & St Louis, MO & 48.22 & 18.00 & 44.80 & 0.58 & 0.35 & 0.04 & 0.16 & 0.38 & 39.70 \\ 
  29 & Tampa, FL & 95.61 & 16.00 & 49.60 & 0.75 & 0.14 & 0.20 & 0.15 & 0.32 & 3.95 \\ 
  30 & Wilmington, DE & 66.21 & 16.00 & 60.27 & 0.63 & 0.26 & 0.10 & 0.14 & 0.30 & 29.60 \\ 
   \hline\hline
\end{tabular}
}
\end{table}

\begin{table}
  \centering
  \caption{Callaway \& Sant'Anna DR-DiD UPDATED 24 MAY}
  \begin{adjustbox}{width=\textwidth}
\begin{tabular}{l c c c c | c c c c}
\hline
 & C1 & C2 & C3 & Benchmark & D1 & D2 & D3 & Benchmark \\
\hline
-8 & $-0.08$          & $0.16$           & $-0.04$          & $0.16$           & $0.06$           & $0.33$           & $-0.28$          & $1.19$           \\
   & $ [-0.59; 0.42]$ & $ [-0.19; 0.52]$ & $ [-0.31; 0.23]$ & $ [-0.38; 0.69]$ & $ [-0.60; 0.73]$ & $ [-0.32; 0.98]$ & $ [-1.42; 0.86]$ & $ [-0.66; 3.05]$ \\
-7 & $0.01$           & $-0.06$          & $0.42^{*}$       & $-0.08$          & $-0.19$          & $-0.18$          & $-0.11$          & $-0.75$          \\
   & $ [-0.36; 0.38]$ & $ [-0.69; 0.57]$ & $ [ 0.00; 0.85]$ & $ [-0.46; 0.30]$ & $ [-0.90; 0.52]$ & $ [-0.92; 0.57]$ & $ [-1.28; 1.07]$ & $ [-2.60; 1.09]$ \\
-6 & $0.21$           & $0.49$           & $0.62$           & $0.26$           & $-0.03$          & $-0.27$          & $-0.10$          & $-0.22$          \\
   & $ [-0.16; 0.58]$ & $ [-0.42; 1.41]$ & $ [-0.42; 1.67]$ & $ [-0.43; 0.96]$ & $ [-0.82; 0.76]$ & $ [-0.82; 0.27]$ & $ [-0.75; 0.56]$ & $ [-0.88; 0.44]$ \\
-5 & $0.20$           & $-0.01$          & $0.24$           & $0.25$           & $-0.02$          & $0.32$           & $0.17$           & $0.23$           \\
   & $ [-0.12; 0.53]$ & $ [-0.71; 0.68]$ & $ [-0.99; 1.46]$ & $ [-0.48; 0.97]$ & $ [-0.59; 0.55]$ & $ [-0.32; 0.97]$ & $ [-0.97; 1.31]$ & $ [-0.43; 0.88]$ \\
-4 & $0.39$           & $0.31$           & $0.52$           & $0.22$           & $0.04$           & $-0.38$          & $-0.32$          & $-0.38$          \\
   & $ [-0.25; 1.02]$ & $ [-0.23; 0.85]$ & $ [-0.25; 1.28]$ & $ [-0.15; 0.60]$ & $ [-0.39; 0.47]$ & $ [-1.00; 0.25]$ & $ [-1.63; 1.00]$ & $ [-1.16; 0.40]$ \\
-3 & $-0.08$          & $0.24$           & $0.51$           & $0.36$           & $-0.07$          & $-0.40$          & $-0.24$          & $0.83$           \\
   & $ [-0.81; 0.64]$ & $ [-0.33; 0.81]$ & $ [-0.57; 1.60]$ & $ [-0.07; 0.80]$ & $ [-0.57; 0.43]$ & $ [-1.47; 0.67]$ & $ [-1.37; 0.90]$ & $ [-0.25; 1.92]$ \\
-2 & $-0.05$          & $-0.07$          & $0.03$           & $-0.06$          & $-0.18$          & $-0.05$          & $-0.05$          & $-0.36$          \\
   & $ [-0.31; 0.21]$ & $ [-0.36; 0.23]$ & $ [-0.44; 0.49]$ & $ [-0.43; 0.32]$ & $ [-0.70; 0.35]$ & $ [-0.49; 0.38]$ & $ [-0.79; 0.68]$ & $ [-1.14; 0.42]$ \\
-1 & $-0.03$          & $0.03$           & $-0.07$          & $-0.16$          & $0.04$           & $0.45$           & $0.62$           & $0.20$           \\
   & $ [-0.35; 0.29]$ & $ [-0.26; 0.32]$ & $ [-0.45; 0.31]$ & $ [-0.61; 0.30]$ & $ [-0.51; 0.60]$ & $ [-0.63; 1.52]$ & $ [-0.54; 1.77]$ & $ [-0.42; 0.82]$ \\
0  & $0.03$           & $-0.07$          & $-0.15$          & $-0.04$          & $0.12$           & $0.32$           & $0.19$           & $0.17$           \\
   & $ [-0.22; 0.29]$ & $ [-0.36; 0.23]$ & $ [-0.52; 0.23]$ & $ [-0.35; 0.28]$ & $ [-0.27; 0.52]$ & $ [-0.39; 1.03]$ & $ [-0.62; 1.00]$ & $ [-0.42; 0.77]$ \\
1  & $0.04$           & $-0.09$          & $-0.21$          & $-0.07$          & $0.07$           & $0.03$           & $-0.12$          & $-0.15$          \\
   & $ [-0.31; 0.40]$ & $ [-0.48; 0.31]$ & $ [-0.82; 0.40]$ & $ [-0.41; 0.27]$ & $ [-0.54; 0.68]$ & $ [-0.70; 0.76]$ & $ [-1.19; 0.95]$ & $ [-0.89; 0.59]$ \\
2  & $0.00$           & $-0.08$          & $-0.29$          & $-0.21$          & $0.08$           & $0.15$           & $0.07$           & $0.21$           \\
   & $ [-0.50; 0.50]$ & $ [-0.63; 0.46]$ & $ [-0.98; 0.40]$ & $ [-0.64; 0.21]$ & $ [-0.61; 0.77]$ & $ [-0.76; 1.06]$ & $ [-1.24; 1.38]$ & $ [-0.52; 0.94]$ \\
3  & $0.13$           & $0.03$           & $-0.08$          & $0.23$           & $0.35$           & $0.37$           & $0.82$           & $0.03$           \\
   & $ [-0.28; 0.54]$ & $ [-0.77; 0.83]$ & $ [-0.93; 0.77]$ & $ [-0.63; 1.09]$ & $ [-0.29; 0.99]$ & $ [-0.25; 0.99]$ & $ [-0.11; 1.75]$ & $ [-0.56; 0.61]$ \\
4  & $0.16$           & $-0.06$          & $-0.17$          & $0.38$           & $0.34$           & $-0.11$          & $-0.16$          & $0.14$           \\
   & $ [-0.44; 0.77]$ & $ [-1.15; 1.02]$ & $ [-1.18; 0.84]$ & $ [-0.59; 1.35]$ & $ [-0.39; 1.06]$ & $ [-2.17; 1.96]$ & $ [-2.49; 2.17]$ & $ [-0.63; 0.91]$ \\
5  & $0.28$           & $0.55$           & $0.39$           & $0.55$           & $-0.02$          & $-0.13$          & $-0.14$          & $-0.15$          \\
   & $ [-0.48; 1.05]$ & $ [-1.20; 2.30]$ & $ [-0.94; 1.72]$ & $ [-0.37; 1.48]$ & $ [-0.84; 0.80]$ & $ [-1.17; 0.90]$ & $ [-1.00; 0.73]$ & $ [-1.17; 0.86]$ \\
6  & $0.64$           & $0.69$           & $0.48$           & $0.79$           & $0.37$           & $0.55$           & $0.49$           & $-0.01$          \\
   & $ [-0.06; 1.35]$ & $ [-1.53; 2.90]$ & $ [-1.09; 2.05]$ & $ [-0.30; 1.88]$ & $ [-0.52; 1.27]$ & $ [-0.71; 1.80]$ & $ [-0.47; 1.45]$ & $ [-0.83; 0.80]$ \\
7  & $0.75$           & $0.64$           & $0.54$           & $0.94$           & $0.49$           & $0.53$           & $0.41$           & $0.36$           \\
   & $ [-0.06; 1.55]$ & $ [-1.50; 2.78]$ & $ [-1.33; 2.40]$ & $ [-0.17; 2.06]$ & $ [-0.42; 1.40]$ & $ [-1.14; 2.20]$ & $ [-0.91; 1.73]$ & $ [-1.03; 1.75]$ \\
8  & $0.61$           & $0.56$           & $0.55$           & $0.96$           & $0.68$           & $1.05$           & $0.85$           & $0.65$           \\
   & $ [-0.04; 1.27]$ & $ [-1.13; 2.25]$ & $ [-1.03; 2.13]$ & $ [-0.15; 2.07]$ & $ [-0.30; 1.65]$ & $ [-0.84; 2.93]$ & $ [-0.65; 2.34]$ & $ [-0.18; 1.49]$ \\
9  & $0.42$           & $0.39$           & $0.33$           & $0.91$           & $0.65$           & $1.03$           & $0.97$           & $0.71$           \\
   & $ [-0.41; 1.25]$ & $ [-1.28; 2.06]$ & $ [-1.09; 1.75]$ & $ [-0.16; 1.99]$ & $ [-0.97; 2.27]$ & $ [-0.56; 2.62]$ & $ [-0.64; 2.57]$ & $ [-0.35; 1.77]$ \\
10 & $-0.07$          & $0.23$           & $0.25$           & $0.89$           & $0.35$           & $0.96^{*}$       & $0.85^{*}$       & $1.22^{*}$       \\
   & $ [-0.66; 0.52]$ & $ [-0.60; 1.07]$ & $ [-0.70; 1.19]$ & $ [-0.08; 1.87]$ & $ [-1.00; 1.69]$ & $ [ 0.35; 1.58]$ & $ [ 0.17; 1.53]$ & $ [ 0.27; 2.17]$ \\
11 & $0.11$           & $0.01$           & $0.10$           & $0.78$           & $0.25$           & $1.19^{*}$       & $0.96$           & $1.27$           \\
   & $ [-0.45; 0.67]$ & $ [-0.96; 0.99]$ & $ [-0.78; 0.97]$ & $ [-0.26; 1.83]$ & $ [-1.07; 1.57]$ & $ [ 0.18; 2.19]$ & $ [-0.25; 2.17]$ & $ [-0.41; 2.95]$ \\
12 & $0.15$           & $0.06$           & $-0.15$          & $0.93$           & $0.23$           & $0.73$           & $0.59$           & $1.19$           \\
   & $ [-0.59; 0.89]$ & $ [-0.79; 0.90]$ & $ [-0.96; 0.66]$ & $ [-0.21; 2.06]$ & $ [-0.91; 1.38]$ & $ [-0.08; 1.54]$ & $ [-0.35; 1.53]$ & $ [-0.06; 2.44]$ \\
   \bottomrule
\end{tabular}
\end{adjustbox}
\label{table:CSAestimates}
{\raggedright \small{This table reports dynamic treatment effects of the lifting of eviction moratoria on increases in COVID-19 cases from April 20 to December 31, 2020. The outcome variable is the arcsine of the number of cases or deaths, therefore any values above zero are interpreted as an increase in the rate of new cases. Timing indicators in the left column show estimated effect in weeks before and after moratoria are lifted (t = 0 is the week of lifting). Model 1 controls for stringency index, stay at home orders, mask mandates, and log population. Model 2 adds demographics and education covariates. Model 3 adds political point difference and eviction numbers. The benchmark model is identical across the DR-DiD and IWES specifications and controls for average eviction filings, average stringency index, population, and demographic variables. Simultaneous confidence intervals are reported below estimates and are constructed using bootstrap (across groups). \par}}
\end{table}

\begin{table}
\begin{center}
\caption{Sun \& Abraham Results UPDATED 12 MAY 2023}
\begin{adjustbox}{width= 0.8\textwidth}
\begin{tabular}{l c c c c | c c c c}
\hline
 & C1 & C2 & C3 & Benchmark & D1 & D2 & D3 & Benchmark \\
\hline
-8 & $0.12$         & $-0.32$  & $-0.42$  & $-0.52$         & $0.45$         & $0.02$   & $-0.07$  & $-0.21$  \\
   & $(0.54)$       & $(0.41)$ & $(0.37)$ & $(0.34)$        & $(0.44)$       & $(0.41)$ & $(0.37)$ & $(0.35)$ \\
-7 & $0.01$         & $-0.42$  & $-0.49$  & $-0.60^{\cdot}$ & $0.22$         & $-0.15$  & $-0.22$  & $-0.33$  \\
   & $(0.46)$       & $(0.34)$ & $(0.32)$ & $(0.30)$        & $(0.44)$       & $(0.39)$ & $(0.36)$ & $(0.34)$ \\
-6 & $0.42$         & $-0.07$  & $-0.15$  & $-0.32$         & $0.48$         & $0.05$   & $-0.03$  & $-0.22$  \\
   & $(0.50)$       & $(0.38)$ & $(0.34)$ & $(0.31)$        & $(0.41)$       & $(0.38)$ & $(0.35)$ & $(0.33)$ \\
-5 & $0.80$         & $0.23$   & $0.16$   & $-0.07$         & $0.49$         & $0.01$   & $-0.05$  & $-0.29$  \\
   & $(0.61)$       & $(0.41)$ & $(0.37)$ & $(0.34)$        & $(0.46)$       & $(0.40)$ & $(0.37)$ & $(0.35)$ \\
-4 & $0.89$         & $0.29$   & $0.25$   & $-0.03$         & $0.31$         & $-0.19$  & $-0.23$  & $-0.45$  \\
   & $(0.66)$       & $(0.39)$ & $(0.35)$ & $(0.31)$        & $(0.48)$       & $(0.38)$ & $(0.35)$ & $(0.33)$ \\
-3 & $0.81$         & $0.26$   & $0.20$   & $-0.03$         & $0.43$         & $-0.06$  & $-0.11$  & $-0.35$  \\
   & $(0.63)$       & $(0.40)$ & $(0.35)$ & $(0.30)$        & $(0.44)$       & $(0.35)$ & $(0.32)$ & $(0.30)$ \\
-2 & $0.84$         & $0.32$   & $0.23$   & $0.01$          & $0.53$         & $0.10$   & $0.02$   & $-0.21$  \\
   & $(0.59)$       & $(0.36)$ & $(0.32)$ & $(0.27)$        & $(0.45)$       & $(0.36)$ & $(0.31)$ & $(0.29)$ \\
0  & $0.88$         & $0.29$   & $0.07$   & $0.03$          & $0.67$         & $0.15$   & $-0.08$  & $-0.11$  \\
   & $(0.63)$       & $(0.38)$ & $(0.30)$ & $(0.28)$        & $(0.45)$       & $(0.34)$ & $(0.28)$ & $(0.28)$ \\
1  & $0.82$         & $0.32$   & $0.10$   & $0.03$          & $0.85^{\cdot}$ & $0.41$   & $0.18$   & $0.12$   \\
   & $(0.56)$       & $(0.35)$ & $(0.29)$ & $(0.30)$        & $(0.43)$       & $(0.34)$ & $(0.28)$ & $(0.29)$ \\
2  & $0.89$         & $0.35$   & $0.17$   & $0.06$          & $1.00^{*}$     & $0.50$   & $0.30$   & $0.18$   \\
   & $(0.59)$       & $(0.37)$ & $(0.29)$ & $(0.28)$        & $(0.41)$       & $(0.35)$ & $(0.30)$ & $(0.30)$ \\
3  & $0.94$         & $0.39$   & $0.19$   & $0.04$          & $0.95^{*}$     & $0.46$   & $0.25$   & $0.09$   \\
   & $(0.62)$       & $(0.39)$ & $(0.31)$ & $(0.30)$        & $(0.44)$       & $(0.32)$ & $(0.27)$ & $(0.26)$ \\
4  & $0.98^{\cdot}$ & $0.50$   & $0.34$   & $0.19$          & $0.65$         & $0.23$   & $0.06$   & $-0.15$  \\
   & $(0.54)$       & $(0.35)$ & $(0.28)$ & $(0.27)$        & $(0.41)$       & $(0.34)$ & $(0.29)$ & $(0.28)$ \\
5  & $0.93^{\cdot}$ & $0.40$   & $0.28$   & $0.15$          & $0.66$         & $0.23$   & $0.10$   & $-0.07$  \\
   & $(0.55)$       & $(0.33)$ & $(0.28)$ & $(0.25)$        & $(0.40)$       & $(0.31)$ & $(0.28)$ & $(0.28)$ \\
6  & $0.90^{\cdot}$ & $0.43$   & $0.30$   & $0.17$          & $0.71^{\cdot}$ & $0.30$   & $0.17$   & $-0.02$  \\
   & $(0.52)$       & $(0.34)$ & $(0.28)$ & $(0.27)$        & $(0.39)$       & $(0.32)$ & $(0.28)$ & $(0.29)$ \\
7  & $0.60$         & $0.20$   & $0.09$   & $-0.04$         & $0.63^{\cdot}$ & $0.28$   & $0.17$   & $0.00$   \\
   & $(0.44)$       & $(0.28)$ & $(0.24)$ & $(0.23)$        & $(0.36)$       & $(0.29)$ & $(0.26)$ & $(0.28)$ \\
8  & $0.65$         & $0.25$   & $0.22$   & $0.02$          & $0.64^{\cdot}$ & $0.29$   & $0.25$   & $0.03$   \\
   & $(0.48)$       & $(0.31)$ & $(0.28)$ & $(0.24)$        & $(0.37)$       & $(0.32)$ & $(0.29)$ & $(0.31)$ \\
9  & $0.50$         & $0.16$   & $0.17$   & $-0.12$         & $0.67^{\cdot}$ & $0.34$   & $0.35$   & $0.04$   \\
   & $(0.45)$       & $(0.30)$ & $(0.28)$ & $(0.25)$        & $(0.36)$       & $(0.28)$ & $(0.27)$ & $(0.26)$ \\
10 & $0.74$         & $0.35$   & $0.36$   & $0.09$          & $0.69^{\cdot}$ & $0.32$   & $0.32$   & $0.03$   \\
   & $(0.52)$       & $(0.35)$ & $(0.33)$ & $(0.27)$        & $(0.38)$       & $(0.31)$ & $(0.29)$ & $(0.27)$ \\
11 & $0.70$         & $0.31$   & $0.25$   & $0.11$          & $0.67^{*}$     & $0.30$   & $0.24$   & $0.07$   \\
   & $(0.45)$       & $(0.31)$ & $(0.28)$ & $(0.28)$        & $(0.32)$       & $(0.26)$ & $(0.24)$ & $(0.25)$ \\
12 & $0.58$         & $0.27$   & $0.21$   & $0.10$          & $0.52^{\cdot}$ & $0.21$   & $0.15$   & $0.00$   \\
   & $(0.41)$       & $(0.29)$ & $(0.26)$ & $(0.25)$        & $(0.30)$       & $(0.25)$ & $(0.24)$ & $(0.25)$ \\
\hline
\end{tabular}
\end{adjustbox}
\label{table:IWESestimates}
\end{center}
{\raggedright \footnotesize{This table reports dynamic treatment effects of the lifting of eviction moratoria on increases in COVID-19 cases from April 20 to December 31, 2020. The outcome variable is $asin(number of cases)$ or $asin(number of deaths)$, therefore any values above zero are interpreted as an increase in the rate of new cases. Timing indicators in the left column show estimated effect in weeks before and after moaratia are lifted (t = 0 is the week of lifting). Model 1 controls for stringency index, stay at home orders, mask mandates, and log population. Model 2 adds demographics and education covaraites. Model 3 adds political point difference and eviction numbers. Standard errors are clustered at the FIPS level. The benchmark model is identical across the DR-DiD and IWES specifications and controls for average eviction filings, average stringency index, population, and demographic variables. $^\cdot$ indicates significance above the 10\% level. $^*$ indicates significance above the 5\% level. \par}}
\end{table}

\begin{figure}[t]
    %\centering
    \caption{Effect of Moratorium End on New Cases/Deaths: Ben-Michael Augmented Synthetic Control, Ind \label{fig:ASC2}}
    \centering     %%% not \center
    \subfigure[Panel A: Cases]{\includegraphics[width=80mm]{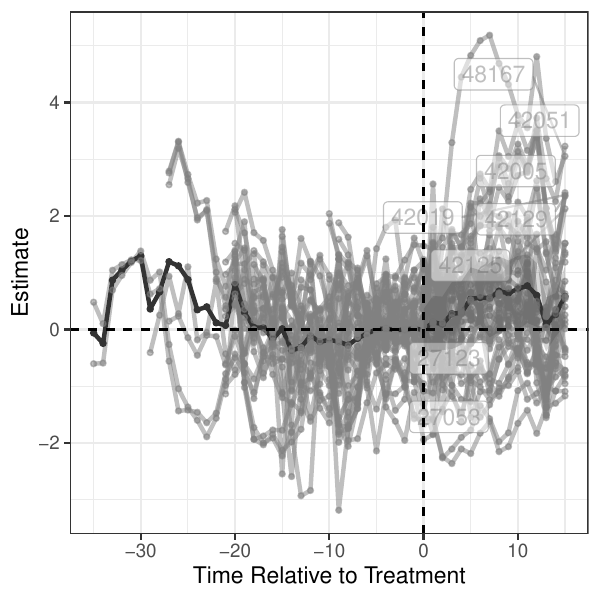}}
    \subfigure[Panel B: Deaths]{\includegraphics[width=80mm]{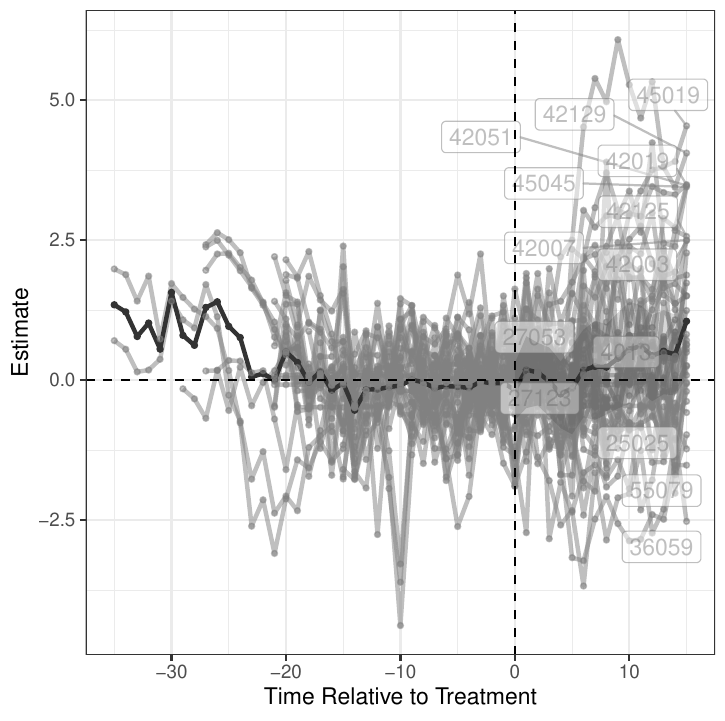}}
{\footnotesize \justifying \singlespacing{This figure depicts pre-treatment balance and individual treatment effects using the Augmented Synthetic Control Method (Ben-Michael, 2021). \textbf{Panel A} shows cases and \textbf{Panel B} depicts deaths. All moratoria are analyzed at the FIPS level. Our outcome variable is the arcsine of the number of COVID-19 deaths, therefore any values above zero are interpreted as an increase in the rate of new deaths.} \par}
\end{figure}

\begin{table}[ht]
\centering
\caption{ASC Results: Cases \& Deaths \label{tab:ASC}}
\adjustbox{max width=\textwidth}{%
\begin{tabular}{rrrrr|rrrrr}
  \hline
  \multicolumn{5}{c}{\textbf{Cases}} & \multicolumn{5}{c}{\textbf{Deaths}}\\
Event Time & Estimate & SE & Upper & Lower & Event Time & Estimate & SE & Upper & Lower \\ 
  \hline
-8 & -0.27 & 0.22 & -0.71 & 0.13 & -8 & -0.04 & 0.34 & -0.74 & 0.49 \\ 
  -7 & -0.15 & 0.20 & -0.58 & 0.21 & -7 & -0.14 & 0.26 & -0.65 & 0.29 \\ 
  -6 & -0.07 & 0.17 & -0.41 & 0.21 & -6 & -0.11 & 0.15 & -0.41 & 0.17 \\ 
  -5 & 0.00 & 0.16 & -0.32 & 0.29 & -5 & -0.12 & 0.19 & -0.53 & 0.23 \\ 
  -4 & 0.02 & 0.14 & -0.27 & 0.25 & -4 & -0.14 & 0.27 & -0.73 & 0.28 \\ 
  -3 & -0.01 & 0.10 & -0.22 & 0.17 & -3 & -0.03 & 0.34 & -0.83 & 0.46 \\ 
  -2 & -0.00 & 0.13 & -0.28 & 0.24 & -2 & -0.06 & 0.26 & -0.63 & 0.35 \\ 
  -1 & 0.02 & 0.17 & -0.33 & 0.33 & -1 & -0.05 & 0.26 & -0.59 & 0.38 \\ 
  0 & -0.01 & 0.18 & -0.36 & 0.36 & 0 & -0.17 & 0.27 & -0.71 & 0.30 \\ 
  1 & 0.13 & 0.25 & -0.41 & 0.59 & 1 & 0.19 & 0.33 & -0.50 & 0.78 \\ 
  2 & 0.09 & 0.24 & -0.36 & 0.58 & 2 & 0.14 & 0.29 & -0.45 & 0.67 \\ 
  3 & 0.28 & 0.31 & -0.34 & 0.84 & 3 & -0.02 & 0.32 & -0.65 & 0.58 \\ 
  4 & 0.29 & 0.29 & -0.27 & 0.85 & 4 & -0.25 & 0.30 & -0.86 & 0.33 \\ 
  5 & 0.54 & 0.28 & -0.00 & 1.10 & 5 & -0.15 & 0.41 & -0.97 & 0.62 \\ 
  6 & $0.56^{*}$ & 0.24 & 0.09 & 1.04 & 6 & 0.18 & 0.40 & -0.66 & 0.90 \\ 
  7 & 0.56 & 0.34 & -0.14 & 1.17 & 7 & 0.24 & 0.46 & -0.71 & 1.06 \\ 
  8 & 0.68 & 0.35 & -0.04 & 1.29 & 8 & 0.23 & 0.41 & -0.61 & 1.00 \\ 
  9 & 0.64 & 0.33 & -0.04 & 1.22 & 9 & 0.39 & 0.43 & -0.51 & 1.20 \\ 
  10 & 0.72* & 0.34 & 0.00 & 1.30 & 10 & 0.55 & 0.46 & -0.42 & 1.40 \\ 
  11 & 0.77* & 0.32 & 0.12 & 1.32 & 11 & 0.61 & 0.40 & -0.21 & 1.34 \\ 
  12 & 0.60* & 0.29 & 0.05 & 1.17 & 12 & 0.44 & 0.40 & -0.38 & 1.16 \\ 
   \hline
\end{tabular}
}
{\raggedright \small{This table reports dynamic treatment effects of the lifting of eviction moratoria on increases in COVID-19 cases from April 20 to December 31, 2020, as estimated by Augmented Synthetic Control. The outcome variable is $asin$(\# cases) or $asin$(\# deaths), therefore any values above zero are interpreted as an increase in the rate of new cases or deaths. \par}}
\end{table}

\end{document}